\documentclass[preprint,aps,12pt,showpacs,tightenlines]{revtex4}
\usepackage{amsmath}
\usepackage{amssymb}
\usepackage{epsfig}
\usepackage{graphicx}
\begin{document}
\preprint{NJNU-TH-07-02}
\newcommand{\beq}{\begin{eqnarray}}
\newcommand{\eeq}{\end{eqnarray}}
\newcommand{\non}{\nonumber\\ }
\newcommand{\etap}{\eta^{(\prime)} }

\newcommand{\psl}{ p \hspace{-1.8truemm}/ }
\newcommand{\nsl}{ n \hspace{-2.2truemm}/ }
\newcommand{\vsl}{ v \hspace{-2.2truemm}/ }
\newcommand{\epsl}{\epsilon \hspace{-1.8truemm}/\,  }

\def \cpl{ Chin. Phys. Lett.  }
\def \ctp{ Commun. Theor. Phys.  }
\def \epjc{ Eur. Phys. J. C }
\def \jpg{  J. Phys. G }
\def \npb{  Nucl. Phys. B }
\def \plb{  Phys. Lett. B }
\def \prd{  Phys. Rev. D }
\def \prl{  Phys. Rev. Lett.  }
\def \pr{  Phys. Rep. }
\def \rmp{  Rev. Mod. Phys. }

\title{$B_{s} \to (\rho, \omega, \phi) \eta^{(\prime)}$ Decays in the Perturbative QCD Approach}
\author{Xin-fen Chen} \email{chenxinfen@163.com}
\author{Dong-qin Guo} \email{medongqin@163.com}
\author{Zhen-jun Xiao} \email{xiaozhenjun@njnu.edu.cn}
\affiliation{Department of Physics and Institute of Theoretical Physics, Nanjing Normal
University, Nanjing, Jiangsu 210097, P.R.China}
\date{\today}
\begin{abstract}
In this paper, we calculate the branching ratios and CP-violating
asymmetries for $B_{s} \to ( \rho^0,\omega, \phi) \eta^{(\prime)}$
decays in the perturbative QCD (pQCD) factorization approach.
Numerically we found that (a) the pQCD predictions for the CP-averaged branching ratios are
$Br(B_{s}\to\rho^0 \eta )\approx 0.07 \times 10^{-6}$ ,
$Br(B_{s}\to\rho^{0}\eta^\prime ) \approx 0.10 \times 10^{-6}$ ,
$Br(B_{s} \to\omega\eta  ) \approx 0.02 \times 10^{-6}$,
$Br(B_{s} \to\omega\eta^\prime  ) \approx 0.13 \times 10^{-6}$,
$Br(B_{s} \to\phi\eta  ) \approx 2.7 \times 10^{-5}$ and
$Br(B_{s} \to\phi\eta^\prime ) \approx 2.0 \times 10^{-5}$;
(b) the gluonic contributions are small in size: less than $3\%$ for
$B \to (\rho, \omega, \phi) \eta$ decays, and about $10\%$ for
$B \to (\rho, \omega, \phi) \eta^\prime$ decays;
and (c) the pQCD predictions for the CP-violating asymmetries of the considered decays are
generally not large in magnitude. The above predictions can be
tested in the forthcoming LHC-b experiments at CERN.
\end{abstract}


\maketitle

\section{Introduction}

The theoretical calculations and experimental measurements of the rare B meson decays play an important
role in testing the standard model (SM), probing CP violation of B meson system and
searching for possible new
physics beyond the SM. At present, about 1000 million  events of B meson pair productions and  decays have been
collected by BaBar and Belle collaborations. In the forthcoming LHC experiments,
a huge amount of B meson events, say around $10^{11} \sim 10^{12}$, are expected,  and therefore
the rare B meson decays with a branching ratio around $10^{-7}$ can be observed with good precision.
Another advantage is that the heavier $B_s$ and $B_c$ mesons and b-baryons, besides the $B_{u}$ and $B_d$,
can also be produced and studied at LHC \cite{lhcb}.

By employing the generalized factorization approach\cite{aag,yhc} or  the QCD factorization (QCDF)
approach \cite{mbg13}, about forty $B_s \to h_1 h_2$ ($h_i$ stand for light pseudo-scalar or vector mesons )
decay modes have been studied in the framework of SM \cite{chenbs99,tseng99,jfs03,mb333}
or in some new physics models beyond the SM \cite{xiaobs01}.
In this paper, we will study the $B_{s}\to\rho^0 \etap$ , $\omega \etap$
and $\phi \etap$ decays in the pQCD factorization approach.
In principle, the physics for the $B_{s}$ two-body hadronic
decays is very similar to that for the $B_{d}$ meson except that the
spectator $\emph{d}$ quark is replaced by the $\emph{s}$  quark.

For $B_{s} \to ( \rho^0, \omega, \phi) \etap$  decays, the $B_{s}$ meson is heavy, setting at rest
and decaying into two light mesons (i.e. $\rho^0$ and $\etap$ ) with large momenta.
Therefore the light final state mesons are moving very fast in the rest frame of $B_{s}$
meson. In this case, the short distance hard process dominates the
decay amplitude. We assume that the soft final state interaction is not important for such decays.
The smallness of FSI effects for B meson decays into two light final state mesons
has been put forward by Bjorken \cite{b89} based on the color transparency argument
\cite{lb80}, and also supported by further renormalization group analysis of soft gluon
exchanges among initial and final state mesons \cite{soft}. With the Sudakov resummation, we
can include the leading double logarithms for all loop diagrams,
in association with the soft contribution.
Unlike the usual factorization approach, the hard part
of the pQCD approach consists of six quarks rather than four. We
thus call it six-quark operators or six-quark effective theory.
Applying the six-quark effective theory to $B_{s}$ meson decays, we
need meson wave functions for the hadronization of quarks into
mesons. All the collinear dynamics are included in the meson wave functions.

This paper is organized as follows. In Sec.~\ref{sec:f-work}, we
give a brief review for the PQCD factorization approach. In
Sec.~\ref{sec:p-c}, we calculate analytically the related Feynman
diagrams and present the  decay amplitudes for the studied decay
modes. In Sec.~\ref{sec:n-d}, we show the numerical results for the
branching ratios and CP asymmetries of$B_{s}\to\rho^0 \etap$ , $\omega \etap$
and $\phi \etap$ decays and comparing them with the results obtained in the other two
methods mentioned above. The summary and some discussions are
included in the final section.

\section{ Theoretical Framework}\label{sec:f-work}

The pQCD factorization approach has been developed and
applied in the non-leptonic $B$ meson decays for some time \cite{lb80,hnl53,luy01,hnl66}.
This approach is based on $k_T$ factorization scheme, where three
energy scales  are involved \cite{hnl53}.
In this approach, the decay amplitude is factorized  into the
convolution of the meson's light-cone wave functions, the hard
scattering kernels and the Wilson coefficients, which stand for the
soft $(\Phi)$, hard(H), and harder(C) dynamics respectively.
The hard dynamics (H) describes the four quark operator and
the spectator quark connected by a hard gluon.
This hard part is characterized by $\sqrt{\Lambda M_{B_{s}}}$, and
can be calculated perturbatively in pQCD approach.
The harder dynamics (C)is from $m_W$ scale to $m_{B_{s}}$ scale described by
renormalization group equation for the four quark operators.
The dynamics below $\sqrt{\Lambda M_{B_{s}}}$
is soft, which is described by the meson wave functions $(\Phi)$.
While the function (H) depends on the processes considered, the wave function is
independent of the specific processes. Using the wave functions determined
from other well measured processes, one can make quantitative predictions
here. Based on this factorization, the decay amplitude can be written as the following
\beq
{\cal  A}(B_{s} \to M_1 M_2)\sim \int\!\! d^4k_1 d^4k_2 d^4k_3\
 \mathrm{Tr} \left [ C(t) \Phi_{B_{s}}(k_1) \Phi_{M_1}(k_2)
 \Phi_{M_2}(k_3) H(k_1,k_2,k_3, t) \right ], \label{eq:con1}
 \eeq
where $k_i$'s are momenta of light quarks included in each mesons,
and $\mathrm{Tr}$ denotes the trace over Dirac and color indices.
$C(t)$ is Wilson coefficient of the four quark operator which results
from the radiative corrections at short distance. The functions $\Phi_M$
and \emph{H} are meson wave functions and the hard part respectively.

For the $B_{s}$ meson decays, since the b quark is
rather heavy we consider the $B_{s}$ meson at rest for simplicity.
It is convenient to use light-cone coordinate $(p^+, p^-, {\bf p}_T)$ to describe the
meson's momenta,
\beq
p^\pm = \frac{1}{\sqrt{2}} (p^0 \pm p^3), \quad  and \quad {\bf p}_T = (p^1, p^2).
\eeq
Using these coordinates the $B_{s}$
meson and the two final state meson momenta can be written as:
\beq
P_{B_{s}} = \frac{M_{B_{s}}}{\sqrt{2}} (1,1,{\bf 0}_T), \quad
P_{\rho} = \frac{M_{B_{s}}}{\sqrt{2}}(1,r_{\rho}^{2},{\bf 0}_T), \quad
P_{\eta} = \frac{M_{B_{s}}}{\sqrt{2}} (0,1-r_{\rho}^{2},{\bf 0}_T),
 \eeq
respectively, where $r_\rho=m_\rho/M_{B_{s}}$ and the light
pseudoscalar meson masses have been neglected (here we just take the decay
$B_s \to \rho^0\eta$ as an example). Putting the light (anti-) quark
momenta in $B_{s}$, $\rho^0$ and $\eta$ mesons as $k_1$, $k_2$,
and $k_3$, respectively, we can choose
\beq
 k_1 = (x_1 P_1^+,0,{\bf  k}_{1T}), \quad
 k_2 = (x_2 P_2^+,0,{\bf k}_{2T}), \quad
 k_3 = (0,x_3 P_3^-,{\bf k}_{3T}).
\eeq
Then the integration over $k_1^-$, $k_2^-$, and $k_3^+$ in eq.(\ref{eq:con1})
will lead to
\beq
{\cal A}(B_{s} & \to & \rho^0\eta) \sim \int\!\! d x_1 d x_2 d x_3 b_1 d
      b_1 b_2 d b_2 b_3 d b_3 \non
      && \quad \cdot \mathrm{Tr} \left [ C(t)
      \Phi_{B_{s}}(x_1,b_1) \Phi_{\rho}(x_2,b_2) \Phi_{\eta}(x_3,
      b_3) H(x_i, b_i, t) S_t(x_i)\, e^{-S(t)} \right ], \label{eq:a2}
\eeq
where $b_i$ is the conjugate space coordinate of $k_{iT}$, and
$t$ is the largest energy scale in function $H(x_i,b_i,t)$, as a
function in terms of $x_i$ and $b_i$. The large logarithms ($\ln m_W/t$)
coming from QCD radiative corrections to four quark
operators are included in the Wilson coefficients $C(t)$. The large
double logarithms ($\ln^2 x_i$) on the longitudinal direction are
summed by the threshold resummation ~\cite{hnl66}, and they lead to
$S_t(x_i)$ which smears the end-point singularities on $x_i$. The
last term, $e^{-S(t)}$, is the Sudakov form factor resulting from
overlap of soft and collinear divergences, which suppresses the soft
dynamics effectively.
Thus it makes the perturbative calculation of the hard part $H$ applicable at intermediate scale,
i.e., $M_{B_{s}}$ scale. We will calculate analytically the function
$H(x_i,b_i,t)$ for our six decays in the first
order in $\alpha_s$ expansion and give the convoluted amplitudes in next section.

For the considered $B_s \to \rho^0\eta^{(\prime)},\omega\eta^{(\prime)}$ and
$\phi\eta^{(\prime)}$ decays, the low energy weak effective
Hamiltonian $H_{eff}$ for the $b \to q$ transition with $q=(d,s)$ can be written as \cite{buras96}:
 \beq
\label{eq:heff}
{\cal H}_{eff} = \frac{G_{F}} {\sqrt{2}} \, \left[ V_{ub} V_{uq}^* \left (C_1(\mu) O_1^u(\mu) +
C_2(\mu) O_2^u(\mu) \right) - V_{tb} V_{tq}^* \, \sum_{i=3}^{10}
C_{i}(\mu) \,O_i(\mu) \right] \; .
\eeq
We specify below the operators in ${\cal H}_{eff}$ for $b \to d$ transition:
\beq
\begin{array}{llllll}
O_1^{u} & = &  \bar d_\alpha\gamma^\mu L u_\beta\cdot \bar
u_\beta\gamma_\mu L b_\alpha\ , &O_2^{u} & = &\bar
d_\alpha\gamma^\mu L u_\alpha\cdot \bar
u_\beta\gamma_\mu L b_\beta\ , \\
O_3 & = & \bar d_\alpha\gamma^\mu L b_\alpha\cdot \sum_{q'}\bar
 q_\beta'\gamma_\mu L q_\beta'\ ,   &
O_4 & = & \bar d_\alpha\gamma^\mu L b_\beta\cdot \sum_{q'}\bar
q_\beta'\gamma_\mu L q_\alpha'\ , \\
O_5 & = & \bar d_\alpha\gamma^\mu L b_\alpha\cdot \sum_{q'}\bar
q_\beta'\gamma_\mu R q_\beta'\ ,   & O_6 & = & \bar
d_\alpha\gamma^\mu L b_\beta\cdot \sum_{q'}\bar
q_\beta'\gamma_\mu R q_\alpha'\ , \\
O_7 & = & \frac{3}{2}\bar d_\alpha\gamma^\mu L b_\alpha\cdot
\sum_{q'}e_{q'}\bar q_\beta'\gamma_\mu R q_\beta'\ ,   & O_8 & = &
\frac{3}{2}\bar d_\alpha\gamma^\mu L b_\beta\cdot
\sum_{q'}e_{q'}\bar q_\beta'\gamma_\mu R q_\alpha'\ , \\
O_9 & = & \frac{3}{2}\bar d_\alpha\gamma^\mu L b_\alpha\cdot
\sum_{q'}e_{q'}\bar q_\beta'\gamma_\mu L q_\beta'\ ,   & O_{10} &
= & \frac{3}{2}\bar d_\alpha\gamma^\mu L b_\beta\cdot
\sum_{q'}e_{q'}\bar q_\beta'\gamma_\mu L q_\alpha'\ ,
\label{eq:operators}
\end{array}
\eeq
where $\alpha$ and $\beta$ are the $SU(3)$ color indices; $L,R=(1\mp \gamma_5)$
are the left- and right-handed projection operators.
The sum over $q'$ runs over the quark fields that are active at the scale $\mu=O(m_b)$.
Since we here work at the leading twist approximation and leading double logarithm summation,
we will also use the leading order (LO) expressions of the Wilson coefficients
$C_i(\mu)$ ($i=1,\ldots,10$), although the
next-to-leading order $C_i(\mu)$ already exist in the literature~\cite{buras96}.
This is the consistent way to cancel the explicit $\mu$ dependence in the theoretical formulae.

For the renormalization group evolution of the Wilson coefficients
from higher scale to lower scale, we use the formulae as given in Ref.~\cite{luy01} directly.
At the high $m_W$ scale, the leading order Wilson coefficients $C_i(M_W)$ are simple and can be found
easily in Ref.~\cite{buras96}. In pQCD approach, the scale `t' may be
larger or smaller than the $m_b$ scale. For the case of $ m_b< t<
m_W$, we evaluate the Wilson coefficients at $t$ scale using leading
logarithm running equations, as given in Eq.(C1) of Ref.~\cite{luy01}.
For the case of $t < m_b$, we then evaluate the Wilson
coefficients at $t$ scale by using $C_i(m_b)$ as input
and the formulae given in Appendix D of Ref.~\cite{luy01}.

For the wave function of the heavy $B_s$ meson, we take
\beq
\Phi_{B_{s}}= \frac{1}{\sqrt{2N_c}} (\psl_{B_{s}}
+M_{B_{s}}) \gamma_5 \phi_{B_{s}} ({\bf k_1}).
\label{eq:bmeson}
\eeq
Here only the contribution of Lorentz structure $\phi_{B_{s}} ({\bf k_1})$ is taken into account, since
the contribution of the second Lorentz structure $\bar{\phi}_{B_{s}}$ is numerically small and
can be neglected. The distribution amplitude $\phi_{B_{s}}$ in Eq.~(\ref{eq:bmeson}) will be given lately
in Eq.~(\ref{eq:phib}).

For $B \to V \etap $ decays, the vector meson $V=(\rho,\phi,\omega)$ is longitudinally polarized.
The relevant longitudinal polarized component of the wave function for $\rho$ meson, for example,
is given as ~\cite{tk07},
\beq
\Phi_{\rho}= \frac{1}{\sqrt{2N_c}} \left\{ \epsl \left[m_\rho \phi_\rho (x)
+ \psl_\rho \phi_\rho^t (x) \right]
+m_\rho \phi_\rho^s (x)\right\},  \label{eq:brhok}
\eeq
where the first term is the leading twist wave function (twist-2), while the
second and third terms are sub-leading twist (twist-3) wave
functions. For the case of $V=\omega$ and $\phi$, their wave functions are the same in structure as
that defined in Eq.~(\ref{eq:brhok}), but with different distribution amplitudes.
One can find the distribution amplitudes $\phi_{\omega,\phi}$ and $\phi_{\omega,\phi}^{t,s}(x)$
in next section.

For $\eta$ and $\eta^\prime$ meson, the wave function for $d\bar{d}$ components of $\eta^{(\prime)}$ meson
are given as \cite{ekou2}
\beq
\Phi_{\eta_{d\bar{d}}}(P,x,\zeta)\equiv
\frac{i \gamma_5}{\sqrt{2N_c}} \left [ \psl \phi_{\eta_{d\bar{d}}}^{A}(x)+m_0^{\eta_{d\bar{d}}}
\phi_{\eta_{d\bar{d}}}^{P}(x)+\zeta m_0^{\eta_{d\bar{d}}} (
\vsl \nsl - v\cdot n)\phi_{\eta_{d\bar{d}}}^{T}(x) \right ],
\label{eq:ddbar}
\eeq
where $P$ and $x$ are the momentum and the momentum fraction of
$\eta_{d\bar{d}}$ respectively, while $\phi_{\eta_{d\bar{d}}}^A$,
$\phi_{\eta_{d\bar{d}}}^P$ and $\phi_{\eta_{d\bar{d}}}^T$ represent
the axial vector, pseudoscalar and tensor components of the wave
function respectively, and will be given explicitly in next section.
Following Ref.~\cite{ekou2}, we here also assume that the wave
function of $\eta_{d\bar{d}}$ is the same as $\pi$ wave function based on SU(3) flavor symmetry.
The parameter $\zeta$ is either $+1$ or $-1$ depending on the
assignment of the momentum fraction $x$.

Before we proceed to do the perturbative calculations, we firstly give a brief discussion about the
$\phi-\omega$ mixing , as well as the $\eta-\eta^\prime$ mixing and the gluonic component of the
$\etap$ mesons.

For the vector $\phi-\omega$ meson system, we choose the ideal mixing scheme between $\phi(1020)$
and $\omega(782)$
\beq
\omega= \frac{1}{\sqrt{2}} \left ( u\bar{u} + d \bar{d}\right ), \qquad
\phi = - s\bar{s},
\eeq
since the current data support this ideal mixing scheme \cite{pdg2006}. The quark contents of $\rho^0$ meson
is chosen as $\rho^0=(-u\bar{u} + d\bar{d})/\sqrt{2}$ \cite{pdg2006}.

For the $\eta-\eta^\prime$ system, there exist two popular mixing basis: the octet-singlet basis
and the quark-flavor basis \cite{fk98,0501072}.
Here we  use  the quark-flavor basis \cite{fk98} and define
\beq
\eta_q=(u\bar{u} + d\bar{d})/\sqrt{2}, \qquad \eta_s=s\bar{s}.
\label{eq:qfbasis}
\eeq
The physical states $\eta$ and $\eta^\prime$ are related to $\eta_q$ and $\eta_s$ through a single mixing
angle $\phi$,
\beq
\left( \begin{array}{c}
\eta\\ \eta^\prime \\ \end{array} \right )
&=& U(\phi) \left( \begin{array}{c}
 \eta_q\\ \eta_s \\ \end{array} \right ) =
  \left( \begin{array}{cc}
 \cos{\phi} & -\sin{\phi} \\
 \sin{\phi} & \cos\phi \end{array} \right )
 \left( \begin{array}{c}  \eta_q\\ \eta_s \\ \end{array} \right ).
 \label{eq:e-ep}
\eeq
The corresponding decay constants $f_q, f_s, f_\eta^{q,s}$ and $f_{\eta^\prime}^{q,s}$
have been defined in Ref.~\cite{fk98} as
\beq
<0|\bar q\gamma^\mu\gamma_5 q|\eta_q(P)>  &=& -\frac{i}{\sqrt2}\,f_q\,P^\mu ,\non
<0|\bar s\gamma^\mu\gamma_5 s|\eta_s(P)>  &=& -i f_s\,P^\mu \;, \label{eq:fqfs}\\
<0|\bar q\gamma^\mu\gamma_5 q|\eta^{(\prime)}(P) >  &=& -\frac{i}{\sqrt2}\,f_{\eta^{(\prime)}}^q\,P^\mu \;,\non
<0|\bar s\gamma^\mu\gamma_5 s|\eta^{(\prime)}(P) >  &=& -i f_{\eta^{(\prime)}}^s\,P^\mu \;,
\label{eq:f11}
\eeq
while the decay constants $f_\eta^{q,s}$ and $f_{\eta^\prime}^{q,s}$ are related to $f_q$ and $f_s$
via the same mixing matrix,
\beq
\left(\begin{array}{cc}
f_\eta^q & f_\eta^s \\
   f_{\eta'}^q & f_{\eta'}^s \\
\end{array} \right)= U(\phi)\left(\begin{array}{cc}
  f_q & 0 \\   0 & f_s \\ \end{array} \right)\;.
\label{eq:f12}
\eeq
The three input parameters $f_q$, $f_s$ and $\phi$ in the quark-flavor basis
have been extracted from various related experiments \cite{fk98,0501072}
\beq
f_q = (1.07\pm 0.02) f_\pi, \quad f_s = (1.34 \pm 0.06) f_\pi,
\quad \phi = 39.3^\circ \pm 1.0^\circ,
\eeq
where $f_\pi=130$ MeV. In the numerical calculations, we will use these mixing parameters
as inputs.

As shown in Eq.~(\ref{eq:e-ep}), the physical states $\eta$ and $\eta^\prime$ are generally
considered as a linear combination of light quark pairs $u\bar{u}, d\bar{d}$ and $s\bar{s}$.
But it should be noted that the $\eta^\prime$ meson may have a gluonic component.
Following Ref.~\cite{li0609}, we also estimated the possible
gluonic contributions to $B \to (\rho,\omega, \phi)\etap$ decays induced by the gluonic
corrections to the $B \to \etap$ transition form factors \cite{li0609} and found that these
corrections to both the branching ratios and CP violating asymmetries are indeed small:
less than $10\%$.

Frankly speaking, on the other hand, we currently still do not know how to calculate
reliably the gluonic contributions to the B meson decays involving $\eta^\prime$ meson as final state
particle. For the studied decay modes in this paper, we firstly consider only the dominant contributions from
the quark contents of $\etap$ meson, and then take the subdominant contribution from the possible
gluonic content of $\etap$ meson as a source of theoretical uncertainties.

\section{Perturbative Calculations}\label{sec:p-c}

In this section, we will calculate the hard part $H(t)$ for the considered decays.
Following the same procedure as being used in Refs.~\cite{liu06,wang06,chen071,xiao0606} and taking
$B \to \rho^0 \etap$ decays as an example, we will calculate and show the analytical results for
all decay amplitudes, and then extend the results of $B \to \rho^0 \etap$ decays to other decay modes
under study.

\begin{figure}[t,b]
\vspace{-3 cm}
\centerline{\epsfxsize=21 cm \epsffile{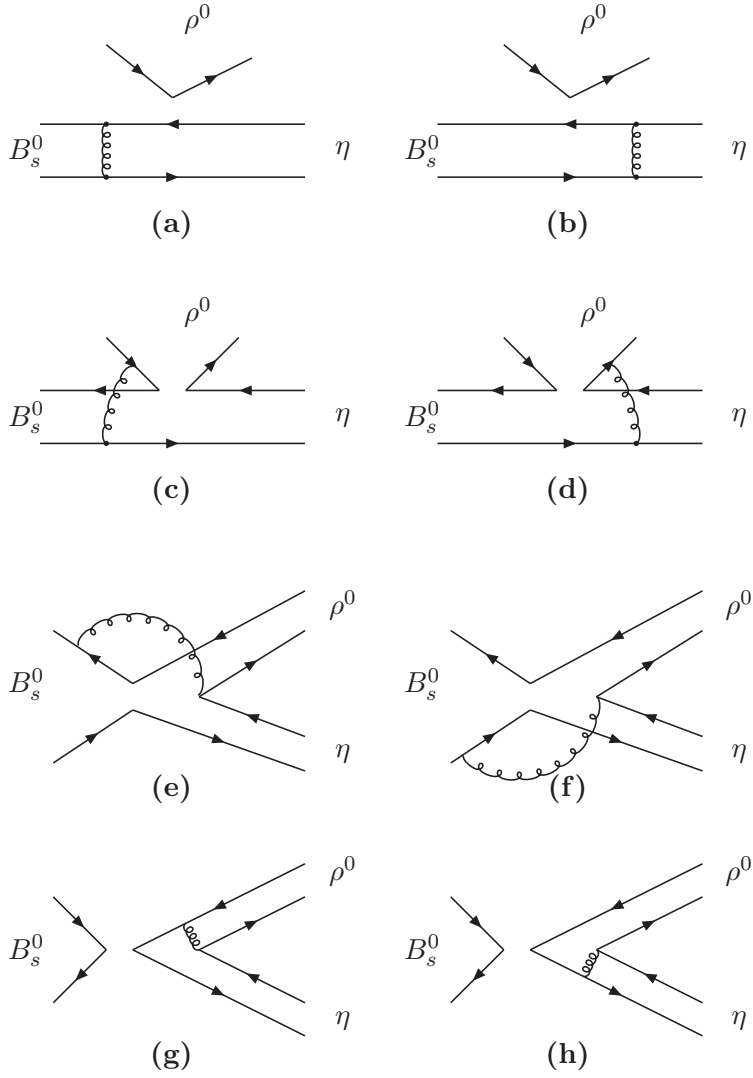}}
\vspace{-14cm}
\caption{ Diagrams contributing to the $B_s\to \rho^0 \eta$ decay (diagram (a) and (b)
contribute to the $B_s\to \eta_{s\overline{s}}$ form factor $F_0^{B_s\to \eta_{s\overline{s}}}$).}
\label{fig:fig1}
\end{figure}

As illustrated in Fig.~\ref{fig:fig1},  there are eight type Feynman diagrams contributing to
the $B_s \to \rho^0 \etap $ decays. We first calculate the usual factorizable diagrams (a) and
(b). Operators $O_1$, $O_2$, $O_3$, $O_4$, $O_9$, and $O_{10}$ are
$(V-A)(V-A)$ currents, the sum of their amplitudes can be written as:
 \beq
 F_{e\eta}&=& -8 \pi C_F
 M_{B_{s}}^{4} f_\rho \int_0^1 d x_{1} dx_{3}\, \int_{0}^{\infty} b_1 db_1
 b_3 db_3\, \phi_{B_{s}}(x_1,b_1)
 \non
  & &
 \cdot \left\{ \left [(1+x_3) \phi_{\eta}^{A} (x_3, b_3)
 +r_{\eta}(1-2x_3)(\phi_{\eta}^{P} (x_3, b_3)+\phi_{\eta}^{T} (x_3,b_3))
 \right] \right.
 \non
  && \left. \alpha_s(t_e^1)
 h_e(x_1,x_3,b_1,b_3)\exp[-S_{ab}(t_e^1)]
 \right.\non
  && \left. +2 r_\eta\phi_{\eta}^{P} (x_3, b_3) \alpha_s(t_e^2)
 h_e(x_3,x_1,b_3,b_1)\exp[-S_{ab}(t_e^2)] \right\} \;.
 \label{eq:ab}
 \eeq
where  $C_F=4/3$ is a color factor and $r_\eta=m_0^{\eta_{s\overline{s}}}/M_{B_s}$ or
$r_\eta=m_0^{\eta_{d\overline{d}}}/M_{B_s}$ since $r_\eta$ depends on the  quark components
in $\eta$. The function $h_e^i$, the scales
$t_e^i$ and the Sudakov factors $S_{ab}$ are displayed in the
appendix. From diagrams Fig.~1(a) and 1(b), one can also extract out the form factor
$F_0^{B_{s}\to \eta_{s\overline{s}} }$.

The operators $O_5$, $O_6$, $O_7$, and $O_8$ have a structure of
$(V-A)(V+A)$. In some decay channels, some of these operators
contribute to the decay amplitude in a factorizable way. Since only
the vector part of $(V{\pm}A)$ current contribute to the scaler
meson production,
\beq
\langle \rho |V+A|B\rangle \langle \eta |V-A | 0 \rangle = \langle \rho
|V-A |B  \rangle \langle \eta |V-A|0 \rangle  ,
 \eeq
the result of these operators is the same as Eq.~(\ref{eq:ab}), and therefore we find easily that
\beq
F_{e\eta}^{P_{1}} &=& F_{e\eta}   \; .
\eeq
In some other cases, one needs to do Fierz transformation  for
these operators to get right flavor and color structure for factorization to
work. In this case, we get $(S+P)(S-P)$ operators from $(V-A)(V+A)$
ones. For these $(S+P)(S-P)$ operators, Fig.~1(a) and 1(b)
will give
\beq
F_{e\eta}^{P_{2}}&=& 0  \; .
\eeq

For the non-factorizable diagrams 1(c) and 1(d), all three meson wave
functions are involved. The integration of $b_3$ can be performed
by using $\delta$ function $\delta(b_3-b_1)$, leaving only integration
of $b_1$ and $b_2$. For the $(V-A)(V-A)$ operators, the corresponding decay amplitude is:
\beq
 M_{e\eta}& = & \frac{32} {\sqrt{6}} \pi C_F M_{B_{s}}^{4}
\int_{0}^{1}d x_{1}d x_{2}\,d x_{3}\,\int_{0}^{\infty} b_1d b_1 b_2d
b_2\,  \non
 & &\cdot
\left.\phi_{B_{s}}(x_1,b_1)\phi_{\rho}(x_2,b_2)\{ x_3
\left[\phi_{\eta}^{A}(x_3,b_1)-2 r_\eta
\phi_{\eta}^{T}(x_3,b_1)\right]\right.\non
  & &\cdot
\left. \alpha_s(t_f)
 h_f(x_1,x_2,x_3,b_1,b_2)\exp[-S_{cd}(t_f)]\right. \} \; .
 \eeq


For the $(S-A)(V+A)$ operators, the decay amplitudes read
\beq
 M_{e\eta}^{P_{1}}& = & \frac{64} {\sqrt{6}} \pi C_F M_{B_{s}}^{4}r_{\rho}
\int_{0}^{1}d x_{1}d x_{2}\,d x_{3}\,\int_{0}^{\infty} b_1d b_1 b_2d
b_2\, \phi_{B_{s}}(x_1,b_1) \non
 & &\cdot\left
\{\left[x_{2}\phi_{\eta}^{A}(x_3,b_1)\left(\phi_{\rho}^{s}(x_2,b_2)-
\phi_{\rho}^{t}(x_2,b_2)\right)+r_{\eta}\left((x_{2}+x_{3})
\left(\phi_{\eta}^{P}(x_3,b_1) \right.\right.\right.\right. \non
 &&\cdot\left.
\phi_{\rho}^{s}(x_2,b_2)+
\phi_{\eta}^{T}(x_3,b_1)\phi_{\rho}^{t}(x_2,b_2)\right)+(x_3-x_2)
(\phi_{\eta}^{P}(x_3,b_1)\phi_{\rho}^{t}(x_2,b_2)
  \non
  &&
\left.\left.\left.+ \phi_{\eta}^{T}(x_3,b_1)\phi_{\rho}^{s}(x_2,b_2)
\right)\right)\right]
 \left.\alpha_s(t_f)
 h_f(x_1,x_2,x_3,b_1,b_2)\exp[-S_{cd}(t_f)]\right. \} , \\
 M_{e\eta}^{P_{2}} & = & - M_{e\eta} \; .
 \eeq

For the non-factorizable annihilation diagrams 1(e) and 1(f), again
all three wave functions are involved. Here we have all three kinds of
contributions. $M_{a\eta}$ is the contribution containing operator type
$(V-A)(V-A)$, while $M_{a\eta}^{P_{1}}$ and $M_{a\eta}^{P_{2}}$ is the
contribution containing operator type $(V-A)(V+A)$ and $(S-P)(S+P)$, respectively
\beq
   M_{a\eta}&=& -\frac{32} {\sqrt{6}} \pi C_F M_{B_{s}}^{4}
 \int_{0}^{1}d x_{1}d x_{2}\,d x_{3}\,\int_{0}^{\infty} b_1d b_1 b_2d
   b_2\, \phi_{B_{s}}(x_1,b_1)\non
  && \cdot
   \left\{ -\left[x_2\phi_{\rho}(x_2,b_2)\phi_{\eta}^{A}(x_3,b_2)+
   r_{\rho}r_{\eta} \left((x_2-x_3)
   \left( \phi_{\eta}^{P}(x_3,b_2)\phi_{\rho}^{t}(x_2,b_2)\right.\right.\right.\right.
  \non
  &&
  \left.+ \phi_{\eta}^{T}(x_3,b_2)\phi_{\rho}^{s}(x_2,b_2)\right)+
   (2+x_2+x_3)\phi_{\eta}^{P}(x_3,b_2)\phi_{\rho}^{s}(x_2,b_2)+(-2+
  \non
  &&
 \left.\left.\left.x_2+x_3)\phi_{\eta}^{T}(x_3,b_2)\phi_{\rho}^{t}(x_2,b_2)\right)\right]
 \alpha_s(t_{f}^{2})h_{f}^{2}(x_1,x_2,x_3,b_1,b_2)\exp[-S_{ef}(t_{f}^{2})]\right.
  \non
  &&
  +\left[x_3\phi_{\rho}(x_2,b_2)\phi_{\eta}^{A}(x_3,b_2)+
  r_{\rho}r_{\eta} \left((x_3-x_2)\left(
  \phi_{\eta}^{P}(x_3,b_2)\phi_{\rho}^{t}(x_2,b_2)+
  \right.\right.\right.
  \non
  &&
  \left.\phi_{\eta}^{T}(x_3,b_2)\phi_{\rho}^{s}(x_2,b_2)\right)+
  (x_2+x_3)\left(\phi_{\eta}^{P}(x_3,b_2)\phi_{\rho}^{s}(x_2,b_2)
  +\phi_{\eta}^{T}(x_3,b_2)\right.
  \non
   &&\cdot\left.\left.\left.
 \left.\phi_{\rho}^{t}(x_2,b_2)\right)\right)\right]\alpha_s(t_{f}^{1})
 h_{f}^{1}(x_1,x_2,x_3,b_1,b_2)\exp[-S_{ef}(t_{f}^{1})]\right. \} \; .
 \eeq
  \beq
   M_{a\eta}^{P_{1}}&=& -\frac{32} {\sqrt{6}} \pi C_F M_{B_{s}}^{4}
 \int_{0}^{1}d x_{1}d x_{2}\,d x_{3}\,\int_{0}^{\infty} b_1d b_1 b_2d
   b_2\, \phi_{B_{s}}(x_1,b_1)\non
  && \cdot
   \left\{ \left[r_{\rho}(2-x_2)\phi_{\eta}^{A}(x_3,b_2)
   \left(\phi_{\rho}^{t}(x_2,b_2)+\phi_{\rho}^{s}(x_2,b_2)\right)
   +r_{\eta}(x_3-2)\phi_{\rho}(x_2,b_2)
   \right.\right.\non
  &&  \cdot
  \left.\left.\left(
  \phi_{\eta}^{P}(x_3,b_2)+\phi_{\eta}^{T}(x_3,b_2)\right)\right]\alpha_s(t_{f}^{2})
  h_{f}^{2}(x_1,x_2,x_3,b_1,b_2)\exp[-S_{ef}(t_{f}^{2})]\right.
  \non
  &&
   +\left[x_2 r_{\rho}\phi_{\eta}^{A}(x_3,b_2)
   \left(\phi_{\rho}^{t}(x_2,b_2)+\phi_{\rho}^{s}(x_2,b_2)\right)
   - x_3 r_{\eta}  \phi_{\rho}(x_2,b_2)
   \right.\non
   &&\cdot
 \left.\left.\left(
  \phi_{\eta}^{P}(x_3,b_2)+\phi_{\eta}^{T}(x_3,b_2)\right)\right]
  \alpha_s(t_{f}^{1})
 h_{f}^{1}(x_1,x_2,x_3,b_1,b_2)\exp[-S_{ef}(t_{f}^{1})]\right. \} \; .
 \eeq
  \beq
   M_{a\eta}^{P_{2}}&=& -\frac{32} {\sqrt{6}} \pi C_F M_{B_{s}}^{4}
 \int_{0}^{1}d x_{1}d x_{2}\,d x_{3}\,\int_{0}^{\infty} b_1d b_1 b_2d
   b_2\, \phi_{B_{s}}(x_1,b_1) \non
  && \cdot
   \left\{ \left[x_3\phi_{\rho}(x_2,b_2)\phi_{\eta}^{A}(x_3,b_2)
   +r_\rho r_\eta \phi_{\eta}^{P}(x_3,b_2) \left((x_2+x_3+2)
   \phi_{\rho}^{s}(x_2,b_2) \right. \right. \right.\non
   &&
   \left.-(x_2-x_3)\phi_{\rho}^{t}(x_2,b_2)\right)+r_\rho r_\eta
   \phi_{\eta}^{T}(x_3,b_2)\left((x_3-x_2)\phi_{\rho}^{s}(x_2,b_2)
   +(x_2\right.\non
  & &
  \left.\left.\left.
  +x_3-2)\phi_{\rho}^{t}(x_2,b_2)\right)\right]
   \alpha_s(t_{f}^{2})
  h_{f}^{2}(x_1,x_2,x_3,b_1,b_2)\exp[-S_{ef}(t_{f}^{2})]\right.\non
  &&
  +\left[(-x_2)\phi_{\rho}(x_2,b_2)\phi_{\eta}^{A}(x_3,b_2)
  +r_\rho r_\eta \left((x_3-x_2)\left(\phi_{\eta}^{P}(x_3,b_2)
  \phi_{\rho}^{t}(x_2,b_2)\right.\right.\right.\non
  && \left.
  +\phi_{\eta}^{T}(x_3,b_2)\phi_{\rho}^{s}(x_2,b_2)\right)
  -(x_2+x_3)\left(\phi_{\eta}^{P}(x_3,b_2)\phi_{\rho}^{s}(x_2,b_2)
  +\phi_{\eta}^{T}(x_3,b_2)\right.\non
  && \left.\left.\left.\left.
  \phi_{\rho}^{t}(x_2,b_2)\right)\right)\right]\alpha_s(t_{f}^{1})
 h_{f}^{1}(x_1,x_2,x_3,b_1,b_2)\exp[-S_{ef}(t_{f}^{1})]\right. \} \; .
  \eeq


The factorizable annihilation diagrams 1(g) and 1(h) involve only
$\rho$ and $\eta$ wave functions.  Again decay amplitude $F_{a\eta}$ is
for $(V-A)(V-A)$ type operators, while the $F_{a\eta}^{P_{1}}$ and
$F_{a\eta}^{P_{2}}$ come from the $(V-A)(V+A)$ and $(S-P)(S+P)$ type operators, respectively
\beq
 F_{a\eta} &=& 8 \pi C_F f_{B_{s}} M_{B_{s}}^{4}\int_{0}^{1}d
 x_{2}\,d x_{3} \,\int_{0}^{\infty} b_2d b_2b_3d b_3 \,  \non
  & &\cdot
  \left\{ \left[ x_3 \phi_\rho(x_2,b_2)
  \phi_{\eta}^{A}(x_3,b_3) + 2 r_\rho r_\eta \phi_{\rho}^{s}(x_2,b_2)
  \left((1+x_3)\phi_{\eta}^{P}(x_3,b_3)\right.\right.\right. \non
   &&
   \left.\left.\left.+(x_3-1)\phi_{\eta}^{T}(x_3,b_3)\right)\right]
   \alpha_s(t_e^3)h_a(x_2,x_3,b_2,b_3)\exp[-S_{gh}(t_e^3)]\right.
   \non
   &&
   \left.-\left[x_2 \phi_\rho(x_2,b_2)
  \phi_{\eta}^{A}(x_3,b_3) + 2 r_\rho r_\eta \phi_{\eta}^{P}(x_3,b_3)
  \left((1+x_2)\phi_{\rho}^{s}(x_2,b_2)\right.\right.\right. \non
   &&
    \left.\left.\left. +(x_2-1)\phi_{\rho}^{t}(x_2,b_2)\right)\right]
   \alpha_s(t_e^4)h_a(x_3,x_2,b_3,b_2)\exp[-S_{gh}(t_e^4)]\right
   \}\;,
   \eeq
   \beq
   F_{a\eta}^{P_{1}}&=&-F_{a\eta}, \\
   F_{a\eta}^{P_{2}} &=& -16 \pi C_F f_{B_{s}} M_{B_{s}}^{4}\int_{0}^{1}d
  x_{2}\,d x_{3} \,\int_{0}^{\infty} b_2d b_2b_3d b_3 \,
  \non
  & &\cdot
   \left\{ \left[2 r_{\rho}\phi_{\rho}^{s}(x_2,b_2)
   \phi_{\eta}^{A}(x_3,b_3)+r_\eta x_3 \phi_{\rho}(x_2,b_2)
   \left( \phi_{\eta}^{P}(x_3,b_3)\right.\right.\right.\non
   &&
    \left.\left.\left.-\phi_{\eta}^{T}(x_3,b_3)\right)\right]
    \alpha_s(t_e^3)h_a(x_2,x_3,b_2,b_3)\exp[-S_{gh}(t_e^3)]\right.
   \non
   &&
   +\left[x_2 r_\rho \phi_{\eta}^{A}(x_3,b_3)\left(\phi_{\rho}^{s}(x_2,b_2)
    -\phi_{\rho}^{t}(x_2,b_2)\right)+2 r_\eta
    \phi_{\rho}(x_2,b_2)\right.\non
   & &\cdot
     \left. \left.\phi_{\eta}^{P}(x_3,b_3)\right]\alpha_s(t_e^4)
    h_a(x_3,x_2,b_3,b_2)\exp[-S_{gh}(t_e^4)]\right \}\;.
    \eeq

When we exchange the position of $\rho$ and $\etap$ mesons in Fig.~\ref{fig:fig1},
then only 4 annihilation diagrams 1(e)-1(h) can contribute to $B \to \rho^0 \etap$ decays.
The corresponding decay amplitudes from  the non-factorizable annihilation diagrams 1(e) and 1(f),
where it is the $\rho^0$ meson who picks up the spectator $s$ quark,  can be written as
\beq
M_{a\rho}&=& -\frac{32} {\sqrt{6}} \pi C_F M_{B_{s}}^{4}
\int_{0}^{1}d x_{1}d x_{2}\,d x_{3}\,\int_{0}^{\infty} b_1d b_1 b_2d
   b_2\, \phi_{B_{s}}(x_1,b_1)\non
     && \cdot
   \left\{ \left[x_3\phi_{\rho}(x_3,b_2)\phi_{\eta}^{A}(x_2,b_2)+
   r_{\rho}r_{\eta} \left((x_3-x_2)
   \left( \phi_{\eta}^{P}(x_2,b_2)\phi_{\rho}^{t}(x_3,b_2)
   \right.\right.\right.\right.\non
     &&
  \left.+ \phi_{\eta}^{T}(x_2,b_2)\phi_{\rho}^{s}(x_3,b_2)\right)+
   (x_2+x_3) \left(\phi_{\eta}^{P}(x_2,b_2)\phi_{\rho}^{s}(x_3,b_2)+
    \phi_{\eta}^{T}(x_2,b_2)\right.\non
    && \cdot\left.\left.\left.
    \phi_{\rho}^{t}(x_3,b_2)\right)\right)\right]\alpha_s(t_{f}^{1})
 h_{f}^{1}(x_1,x_2,x_3,b_1,b_2)\exp[-S_{ef}(t_{f}^{1})]
 -\left[x_2\phi_{\rho}(x_3,b_2) \right.\non
    &&\cdot
   \phi_{\eta}^{A}(x_2,b_2)+
  r_{\rho}r_{\eta} \left((x_2-x_3)\left(
  \phi_{\eta}^{P}(x_2,b_2)\phi_{\rho}^{t}(x_3,b_2)+\phi_{\eta}^{T}(x_2,b_2)
  \right.\right.\non
    &&\cdot\left.\left.
    \phi_{\rho}^{s}(x_3,b_2)\right)\right)+
   r_{\rho}r_{\eta}\left((2+x_2+x_3)\phi_{\eta}^{P}(x_2,b_2)
   \phi_{\rho}^{s}(x_3,b_2)-(2-x_2-x_3)\right.\non
   &&\cdot\left.\left.\left.
   \phi_{\eta}^{T}(x_2,b_2)\phi_{\rho}^{t}(x_3,b_2)
   \right)\right]\alpha_s(t_{f}^{2})
   h_{f}^{2}(x_1,x_2,x_3,b_1,b_2)\exp[-S_{ef}(t_{f}^{2})]\right.\} \; ,
   \eeq
   \beq
   M_{a\rho}^{P_{1}}&=& \frac{32} {\sqrt{6}} \pi C_F M_{B_{s}}^{4}
 \int_{0}^{1}d x_{1}d x_{2}\,d x_{3}\,\int_{0}^{\infty} b_1d b_1 b_2d
   b_2\, \phi_{B_{s}}(x_1,b_1)\non
  && \cdot
   \left\{ \left[x_2 r_\eta \phi_{\rho}(x_3,b_2)\left(
   \phi_{\eta}^{P}(x_2,b_2)+\phi_{\eta}^{T}(x_2,b_2)\right)
   -x_3 r_\rho \left(\phi_{\rho}^{s}(x_3,b_2)\right.\right.\right.
   \non
    && \left.\left.
  +\phi_{\rho}^{t}(x_3,b_2)\right)\phi_{\eta}^{A}(x_2,b_2)
  \right]\alpha_s(t_{f}^{1})
 h_{f}^{1}(x_1,x_2,x_3,b_1,b_2)\exp[-S_{ef}(t_{f}^{1})]\non
  &&
  +\left[(2-x_2)r_\eta \phi_{\rho}(x_3,b_2)\left(
  \phi_{\eta}^{P}(x_2,b_2)+\phi_{\eta}^{T}(x_2,b_2)\right)-(2-x_3)
  r_\rho \left( \phi_{\rho}^{s}(x_3,b_2)\right.\right.
  \non
  &&\left.\left.\left.
  +\phi_{\rho}^{t}(x_3,b_2)\right)
  \phi_{\eta}^{A}(x_2,b_2)\right]\alpha_s(t_{f}^{2})
   h_{f}^{2}(x_1,x_2,x_3,b_1,b_2)\exp[-S_{ef}(t_{f}^{2})]\right.\} \; ,
   \eeq
    \beq
   M_{a\rho}^{P_{2}}&=& \frac{32} {\sqrt{6}} \pi C_F M_{B_{s}}^{4}
 \int_{0}^{1}d x_{1}d x_{2}\,d x_{3}\,\int_{0}^{\infty} b_1d b_1 b_2d
   b_2\, \phi_{B_{s}}(x_1,b_1)\non
  && \cdot
   \left\{ \left[x_2 \phi_{\rho}(x_3,b_2) \phi_{\eta}^{A}(x_2,b_2)
   +r_\rho r_\eta \left((x_2+x_3)\left(\phi_{\eta}^{P}(x_2,b_2)
   \phi_{\rho}^{s}(x_3,b_2)\right.\right.\right.\right.\non
  && \left.
  +\phi_{\eta}^{T}(x_2,b_2)\phi_{\rho}^{t}(x_3,b_2)\right)
  +(x_2-x_3)\left(\phi_{\eta}^{P}(x_2,b_2)\phi_{\rho}^{t}(x_3,b_2)
  +\phi_{\eta}^{T}(x_2,b_2)\right.\non
  && \cdot\left.\left.\left.
  \phi_{\rho}^{s}(x_3,b_2)\right)\right)\right]\alpha_s(t_{f}^{1})
  h_{f}^{1}(x_1,x_2,x_3,b_1,b_2)\exp[-S_{ef}(t_{f}^{1})]
  -\left[x_3 \phi_{\rho}(x_3,b_2)\right.\non
   &&\cdot \left.\left.
    \phi_{\eta}^{A}(x_2,b_2)
   +r_\rho r_\eta \left((x_3-x_2)\left(\phi_{\eta}^{P}(x_2,b_2)
   \phi_{\rho}^{t}(x_3,b_2)+\phi_{\eta}^{T}(x_2,b_2)
   \phi_{\rho}^{s}(x_3,b_2)\right)\right)
   \right.\right.\non
   &&
   +r_\rho r_\eta \left((2+x_2+x_3)\phi_{\eta}^{P}(x_2,b_2)
   \phi_{\rho}^{s}(x_3,b_2)+(x_2+x_3-2)\phi_{\eta}^{T}(x_2,b_2)
   \right.\non
   &&\cdot\left.\left.\left.
   \phi_{\rho}^{t}(x_3,b_2)
   \right)\right]\alpha_s(t_{f}^{2})
   h_{f}^{2}(x_1,x_2,x_3,b_1,b_2)\exp[-S_{ef}(t_{f}^{2})]\right.\} \; .
   \eeq

For the factorizable annihilation diagrams 1(g) and 1(h) after the exchange of $\rho^0$ and $\etap$ mesons,
the corresponding decay amplitudes can be obtained directly through the links with
their counterparts $F_{a\eta}$, $F_{a\eta}$ and $F_{a\eta}^{P_2}$
\beq
F_{a\rho}=-F_{a\eta},\quad F_{a\rho}^{P_1}=F_{a\eta},\quad
F_{a\rho}^{P_2}=F_{a\eta}^{P_2}.
\eeq

Combining the contributions from different diagrams, the total decay
amplitude for $B_{s} \to \rho^0 \eta$ decay can be written as:
\beq
{\cal M}(\rho^0 \eta)&=&  F_{e\eta} \left[\xi_u
  \left( C_1+\frac{1}{3}C_2\right) -\xi_t\left(
  \frac{3}{2}C_7+\frac{1}{2}C_8+\frac{3}{2}C_9+\frac{1}{2}C_{10}\right)\right]
  F_2(\phi)\non
   & &
   +M_{e\eta}\left[\xi_u C_2
  -\xi_t\left(-\frac{3}{2}C_8+\frac{3}{2}C_{10}\right)\right] F_2(\phi)\non
   & &
   +\left(M_{a\eta}+M_{a\rho}\right)\left[\xi_u
   C_2-\xi_t \; \frac{3}{2}C_{10} \right]F_1(\phi)
   +\left( M_{a\eta}^{P_{2}}+M_{a\rho}^{P_{2}}\right)
   \left[-\xi_t\; \frac{3}{2}C_8\right]   F_1(\phi)\non
&&  +\left(F_{a\eta}+F_{a\rho}\right)
   \left[\xi_u\left( C_1+\frac{1}{3}C_2\right)\right.\non
   && \left. -\xi_t\left(-\frac{3}{2}C_7-\frac{1}{2}C_8+\frac{3}{2}C_9
   +\frac{1}{2}C_{10}\right)\right]F_1(\phi), \ \
   \label{eq:m1}
\eeq
where $\xi_u = V_{ub}^*V_{ud}$, $\xi_t = V_{tb}^*V_{td}$, and
 $F_1(\phi)=\cos \phi/2$ and $F_2(\phi)=-\sin \phi/\sqrt{2}$ are the mixing factors.
The Wilson coefficients $C_i$ should be calculated at the appropriate scale $t$ by using
the formulas as given in the Appendices of Ref.~\cite{cdl09}.

Following the same steps, one can derive the total decay
amplitude for $B_{s} \to \omega \eta$ decay:
\beq
{\cal M}(\omega \eta)&=&  F_{e\eta}  \left[\xi_u\left(C_1+\frac{1}{3}C_2\right)
    -\xi_t\left(2 C_3+\frac{2}{3}C_4+2 C_5\right.\right.\non
    &&\left.\left.
    +\frac{2}{3}C_6+\frac{1}{2}C_7+\frac{1}{6}C_8
    +\frac{1}{2}C_9+\frac{1}{6}C_{10}
    \right)\right] F_2(\phi)\non
    & &
    +M_{e\eta}\left[\xi_u \; C_2 -\xi_t\left(2 C_4-2 C_6-\frac{1}{2}C_8+\frac{1}{2}C_{10}
    \right)\right] F_2(\phi)\non
    & &
    +\left(M_{a\eta}+M_{a\omega}\right)
    \left[\xi_u \; C_2
    -\xi_t\left(2 C_4+\frac{1}{2}C_{10}\right)\right] F_1(\phi)\non
    & &
    +\left(M_{a\eta}^{P_{2}}+M_{a\omega}^{P_{2}}\right)
    \left[-\xi_t\left(2C_6+\frac{1}{2}C_8\right)\right]     F_1(\phi)
    +\left( F_{a\eta}+F_{a\omega}\right)
    \left[\xi_u \left(C_1+\frac{1}{3}C_2 \right) \right. \non
    && \left. -\xi_t\left( 2 C_3+\frac{2}{3}C_4-2C_5-\frac{2}{3}C_6
    -\frac{1}{2}C_7-\frac{1}{6}C_8+\frac{1}{2}
    C_9+\frac{1}{6}C_{10}\right)\right]F_1(\phi).
    \label{eq:m2}
\eeq
The individual decay amplitudes $(F_{e\eta}, M_{e\eta}, \cdots)$ in Eq.~(\ref{eq:m2})
can be obtained easily from those as given for the case of $B_s \to \rho^0 \eta$ decay
by the simple replacements
\beq
f_\rho\;\longrightarrow \;f_\omega ,\quad
    f_\rho^T\;\longrightarrow \;f_\omega^T ,\quad
    m_\rho\;\longrightarrow \;m_\omega
\eeq
Note that the difference in the quark components of the two vector mesons $\rho^0$ and $\omega$
has been taken into account.

For $B_s \to \phi \etap$ decays, all eight Feynman diagrams when the $\etap$ or $\phi$ meson picks
up the spectator $s$ quark. For the later case, we firstly consider the factorizable diagrams 1(a) and 1(b).
The decay amplitude $F_{e\phi}$ induced by inserting the $(V-A)(V+A)$ operators is
\beq
 F_{e\phi}&=& -8 \pi C_F  M_{B_{s}}^{4} \int_0^1 d x_{1} dx_{3}\, \int_{0}^{\infty} b_1 db_1
 b_3 db_3\, \phi_{B_{s}}(x_1,b_1)
  \non
  & &
 \cdot \left\{ \left [(1+x_3) \phi_\phi(x_3, b_3)
 +r_{\phi}(1-2x_3)(\phi_{\phi}^{s} (x_3, b_3)+\phi_{\phi}^{t} (x_3,b_3))
 \right] \right.
 \non
  & &
 \left. \alpha_s(t_e^1) h_e(x_1,x_3,b_1,b_3)\exp[-S_{ab}(t_e^1)]
 \right.\non
  & &
  \left. +2 r_\phi \phi_{\phi}^{s} (x_3, b_3) \alpha_s(t_e^2)
 h_e(x_3,x_1,b_3,b_1)\exp[-S_{ab}(t_e^2)] \right\} \;,
 \label{eq:fephi}
  \eeq
where $r_\phi=m_\phi/M_{B_s}$. One can also extract out the $B_s \to \phi$ form factor
$\emph{A}_0^{B_s \to \phi}$ from the diagrams 1(a) and 1(b) when $\phi$ meson picks up the spectator $s$
quark. For other diagrams, the relevant decay amplitudes can be written as
\beq
F_{e\phi}^{P_1}&=&-F_{e\phi}, \\
F_{e\phi}^{P_2}&=& -16 \pi C_F M_{B_{s}}^{4}  r_\eta \int_0^1 d x_{1} dx_{3}\, \int_{0}^{\infty} b_1 db_1
 b_3 db_3\, \phi_{B_{s}}(x_1,b_1)  \non
   & &\cdot
   \left\{ \left [\phi_\phi(x_3, b_3)+r_\phi \left((x_3+2)
   \phi_{\phi}^{s} (x_3, b_3)-x_3 \phi_{\phi}^{t} (x_3, b_3)
   \right)\right]\right.   \non
   &&\cdot
   \alpha_s(t_e^1) h_e(x_1,x_3,b_1,b_3)\exp[-S_{ab}(t_e^1)]
   +\left(x_1 \phi_\phi(x_3, b_3)\right.
   \non
   && \left.\left.
   +2r_\phi \phi_{\phi}^{s} (x_3, b_3)\right)\alpha_s(t_e^2)
 h_e(x_3,x_1,b_3,b_1)\exp[-S_{ab}(t_e^2)] \right\} \;,
\eeq
\beq
  M_{e\phi}& = & \frac{32} {\sqrt{6}} \pi C_F M_{B_{s}}^{4}
\int_{0}^{1}d x_{1}d x_{2}\,d x_{3}\,\int_{0}^{\infty} b_1d b_1 b_2d b_2\,
  \non
  &&\cdot \phi_{B_{s}}(x_1,b_1)
  \phi_{\eta}^{A}(x_2,b_2)\left\{ x_3 \left[\phi_{\phi}(x_3,b_1)-2
  r_\phi \phi_{\phi}^{t}(x_3,b_1)\right]\right.\non
  && \cdot
  \left. \alpha_s(t_f) h_f(x_1,x_2,x_3,b_1,b_2)\exp[-S_{cd}(t_f)]\right. \} \;, \\
 M_{e\phi}^{P_1}&=& 0, \quad M_{e\phi}^{P_2}=M_{e\phi}.
  \eeq

Finally, the total decay amplitude of $B_s \to \phi \eta$ decay can be written as
 \beq
    {\cal M}(\phi \eta) &=& \sqrt{2} F_{e\phi}\left\{    \left[\xi_u\left(C_1+\frac{1}{3}C_2\right)
     \right.\right.\non
    & &
    \left.\left.-\xi_t\left(2C_3+\frac{2}{3}C_4-2C_5-\frac{2}{3}C_6
    -\frac{1}{2}C_7-\frac{1}{6}C_8+\frac{1}{2}C_9
    +\frac{1}{6}C_{10}\right)\right]f_qF_1(\phi)\right.\non
    & &\left.
    -\xi_t\left(\frac{4}{3}C_3+\frac{4}{3}C_4-C_5-\frac{1}{3}C_6
    +\frac{1}{2}C_7+\frac{1}{6}C_8-\frac{2}{3}C_9-\frac{2}{3}C_{10}
    \right)f_sF_2(\phi) \right\}\non
    & &
    + \sqrt{2}F_{e\phi}^{P_2}\left[-\xi_t\left(\frac{1}{3}C_5+C_6
    -\frac{1}{6}C_7-\frac{1}{2}C_8 \right)\right]f_sF_2(\phi)\non
    & &
    +\sqrt{2} F_{e\eta}\left[-\xi_t\left(\frac{4}{3}C_3+\frac{4}{3}C_4+C_5
    +\frac{1}{3}C_6 -\frac{1}{2}C_7-\frac{1}{6}C_8-\frac{2}{3}C_9-
    \frac{2}{3}C_{10}\right)\right]F_2(\phi)\non
    & &
    +\sqrt{2} M_{e\phi}\left\{\left[\xi_u\left(C_2\right)-\xi_t
    \left(2C_4+2C_6+\frac{1}{2}C_8+\frac{1}{2}C_{10}\right)\right]
    F_1(\phi)\right.\non
    & &
    \left.-\xi_t\left(C_3+C_4+C_6-\frac{1}{2}C_8-\frac{1}{2}C_9
    -\frac{1}{2}C_{10}\right)F_2(\phi)\right\}\non
    & &
    + \sqrt{2}M_{e\eta}\left[-\xi_t\left(C_3+C_4-C_6+\frac{1}{2}C_8
    -\frac{1}{2}C_9-\frac{1}{2}C_{10}\right)\right]F_2(\phi)\non
    & &
    + \sqrt{2} M_{e\eta}^{P_{1}}\left[-\xi_t\left(C_5-\frac{1}{2}C_7
    \right)\right]F_2(\phi)\non
    & &
    + \sqrt{2} \left(M_{a\eta}+M_{a\phi}\right)
    \left[-\xi_t\left(C_3+C_4-\frac{1}{2}C_9
    -\frac{1}{2}C_{10}\right)\right]F_2(\phi)\non
    & &
    +\sqrt{2} \left(M_{a\eta}^{P_{1}}+M_{a\phi}^{P_{1}}\right)
    \left[-\xi_t\left(C_5-\frac{1}{2}C_7
    \right)\right]F_2(\phi)\non
    & &
    +\sqrt{2} \left(M_{a\eta}^{P_{2}}+M_{a\phi}^{P_{2}}\right)
    \left[-\xi_t\left(C_6-\frac{1}{2}C_8
    \right)\right]F_2(\phi)\non
    & &
    +\sqrt{2} \left(F_{a\eta}+F_{a\phi}\right)
    \left[-\xi_t\left(\frac{4}{3}C_3
    +\frac{4}{3}C_4-C_5\right.\right.\non
    & & \left.\left.
    -\frac{1}{3}C_6+\frac{1}{2}C_7
    +\frac{1}{6}C_8-\frac{2}{3}C_9-\frac{2}{3}C_{10}
   \right)\right]F_2(\phi)\non
    & &
    +\sqrt{2} \left(F_{a\eta}^{P_{2}}+F_{a\phi}^{P_{2}}\right)
    \left[-\xi_t\left(\frac{1}{3}C_5
    +C_6-\frac{1}{6}C_7-\frac{1}{2}C_8
    \right)\right]F_2(\phi)  .\label{eq:m3}
    \eeq
The individual decay amplitudes $(F_{e\eta}, M_{e\eta}, \cdots)$ in Eq.~(\ref{eq:m3})
can be obtained easily from those as given for the case of $B_s \to \rho^0 \eta$ decay
by the simple replacements
\beq
    f_\rho\;\longrightarrow \;f_\phi ,\quad
    f_\rho^T\;\longrightarrow \;f_\phi^T ,\quad
    m_\rho\;\longrightarrow \;m_\phi \non
    \phi_\rho\;\longrightarrow \;\phi_\phi,\quad
    \phi_\rho^t\;\longrightarrow \;\phi_\phi^t,\quad
    \phi_\rho^s\;\longrightarrow \;\phi_\phi^s
\eeq

The total decay amplitudes for $B_s \to \rho^0 \eta^{\prime}$,
 $B_s \to \omega \eta^{\prime}$ and $B_s \to \phi \eta^{\prime}$
can be obtained easily from Eqs.(\ref{eq:m1}), (\ref{eq:m2}) and (\ref{eq:m3})
by the following simple replacements:
 \beq
F_1(\phi)=\cos{\phi} &\longrightarrow & F'_1(\phi)= \sin{\phi}, \non
F_2(\phi)=-\sin\phi/\sqrt{2} &\longrightarrow & F'_2(\phi) = \cos{\phi}/\sqrt{2}.
 \eeq
Note that the possible gluonic component of $\eta'$ meson has been neglected here. We will estimate
its effects in next section.

\section{Numerical results and Discussions}\label{sec:n-d}

\subsection{Input parameters and wave functions}

We use the following input parameters in the numerical calculations
 \beq
  \Lambda_{\overline{\mathrm{MS}}}^{(f=4)} &=& 250 {\rm MeV}, \quad
  f_\rho = 205 {\rm MeV},\quad f_\rho^T = 160 {\rm MeV}, \non
  m_0^{\eta_{d\overline{d}}} &=& 1.4 {\rm GeV}, \quad f_{B_{s}} = 230 {\rm MeV},
  \quad f_\pi = 130  {\rm MeV}, \non
  m_\omega &=& 0.782 {\rm GeV},\quad f_\omega = 195 {\rm MeV},
  \quad f_\omega^T = 140 {\rm MeV}, \non
  m_\phi &=& 1.02 {\rm GeV},\quad f_\phi = 237 {\rm MeV},
  \quad f_\phi^T = 220 {\rm MeV}, \non
  m_\rho &=& 0.770 {\rm GeV},\quad M_{B_{s}} = 5.37 {\rm GeV},
  \quad M_W = 80.42 {\rm GeV} .
  \label{para}
  \eeq

For the Cabibbo-Kobayashi-Maskawa (CKM) matrix elements, here we adopt the Wolfenstein
parametrization for the CKM matrix up to $\mathcal{O}(\lambda^5)$ with the parameters
$\lambda=0.2272, A=0.818, \bar{\rho}=0.221$ and $\bar{\eta}=0.340$\cite{pdg2006}.

$B_{s}$ meson is different from B meson due to the heavier
strange quark (compare to \emph{u},\emph{d} quark) which induces
the SU(3) symmetry-breaking effect. This effect is
considered to be small and the distribution amplitude of
$B_{s}$ meson (given in the following formula) should be  similar to that of the B meson,
  \beq
 \phi_{B_{s}}(x,b) &=&  N_{B_{s}} x^2(1-x)^2 \mathrm{exp} \left
 [ -\frac{M_{B_{s}}^{2}\ x^2}{2 \omega_{B_s}^2} -\frac{1}{2} (\omega_{B_s} b)^2\right],
 \label{eq:phib}
 \eeq
where $\omega_{B_s}$ is a free parameter in nature. After considering the constraints from
some $B_{s}$ non-leptonic decays \cite{mr04}, we here use $\omega_{B_s}=0.55 \pm 0.05$ in the
numerical calculation.

For the light meson wave function, we neglect the $b$ dependant part, which is not important in
numerical analysis. We choose the wave function of $\rho(\omega$) meson similar to the pion case
~\cite{pb23}:
  \beq
  \phi_{\rho(\omega)}(x) &=& \frac{3}{\sqrt{6} }
  f_{\rho(\omega)}  x (1-x)  \left[1+ 0.18C_2^{\frac{3}{2 }} (2x-1) \right],\\
    \phi_{\rho(\omega)}^t(x) &=&  \frac{f_{\rho(\omega)}^T }{2\sqrt{6} }
  \left\{  3 (2 x-1)^2 +0.3(2 x-1)^2  \left[5(2 x-1)^2-3  \right]
  \right.\nonumber\\
  &&~~\left. +0.21 [3- 30 (2 x-1)^2 +35 (2 x-1)^4] \right\},\\
  \phi_{\rho(\omega)}^s(x) &=&  \frac{3}{2\sqrt{6} }
  f_{\rho(\omega)}^T   (1-2x)  \left[1+ 0.76 (10 x^2 -10 x +1) \right] .
 \eeq

For the wave function of $\phi$ meson, we use ~\cite{pb23}:
   \beq
   \phi_{\phi}(x) &=& \frac{3}{\sqrt{6} }
   f_\phi x (1-x) , \\
   \phi_{\phi}^t(x) &=&  \frac{f_{\phi}^T }{2\sqrt{6} }
   \left\{  3 (1-2 x)^2 +1.68C_4^{\frac{1}{2 }} (1-2x) +
   0.69\left[1+(1-2x)\ln\frac{x}{1-x}\right]\right\} ,\\
   \phi_{\phi}^s(x) &=&  \frac{f_{\phi}^T }{4\sqrt{6} }
   \left[3(1-2x)(4.5-11.2x+11.2 x^2)+1.38\ln\frac{x}{1-x}\right],
   \eeq
where the Gegenbauer polynomials are defined by
 \beq
 C_2^{\frac{3}{2 }} (\xi) &=& \frac{3}{2} \left (5\xi^2-1 \right ),\\
 C_4^{\frac{1}{2 }} (\xi)&=& \frac{1}{8} \left(35\xi^4-
 30\xi^2+3\right) .
 \eeq

For $\eta$ meson's wave function, $\phi_{\eta_{d\bar{d}}}^A$,
$\phi_{\eta_{d\bar{d}}}^P$ and $\phi_{\eta_{d\bar{d}}}^T$ represent
the axial vector, pseudoscalar and tensor components of the wave
function respectively, for which we utilize the results
from the light-cone sum rule~\cite{pb05} including twist-3 contribution:
\beq
  \phi_{\eta_{d\bar{d}}}^A(x)&=&\frac{3}{\sqrt{2N_c}}f_qx(1-x)
  \left\{ 1+a_2^{\eta_{d\bar{d}}}\frac{3}{2}\left [5(1-2x)^2-1 \right
  ]\right. \non &&\left. + a_4^{\eta_{d\bar{d}}}\frac{15}{8} \left
  [21(1-2x)^4-14(1-2x)^2+1 \right ]\right \},  \non
  \phi^P_{\eta_{d\bar{d}}}(x)&=&\frac{1}{2\sqrt{2N_c}}f_q \left \{ 1+
  \frac{1}{2}\left (30\eta_3-\frac{5}{2}\rho^2_{\eta_{d\bar{d}}}
  \right ) \left [ 3(1-2x)^2-1 \right] \right.  \non && \left. +
  \frac{1}{8}\left
  (-3\eta_3\omega_3-\frac{27}{20}\rho^2_{\eta_{d\bar{d}}}-
  \frac{81}{10}\rho^2_{\eta_{d\bar{d}}}a_2^{\eta_{d\bar{d}}} \right )
  \left [ 35 (1-2x)^4-30(1-2x)^2+3 \right ] \right\} ,  \non
  \phi^T_{\eta_{d\bar{d}}}(x) &=&\frac{3}{\sqrt{2N_c}}f_q(1-2x) \non
  && \cdot \left [ \frac{1}{6}+(5\eta_3-\frac{1}{2}\eta_3\omega_3-
  \frac{7}{20}\rho_{\eta_{d\bar{d}}}^2
  -\frac{3}{5}\rho^2_{\eta_{d\bar{d}}}a_2^{\eta_{d\bar{d}}})(10x^2-10x+1)\right  ],
\eeq
with the updated Gegenbauer moments\cite{pball}
    \beq
   a^{\eta_{d\bar{d}}}_2&=& 0.115, \quad    a^{\eta_{d\bar{d}}}_4=-0.015,\quad
   \non
   \rho_{\eta_{d\bar{d}}}&=&m_{\pi}/{m_0^{\eta_{d\bar{d}}}}, \quad
   \eta_3=0.015~, \quad \omega_3=-3.0
    \eeq

 We assume that the wave function of $u\bar{u}$ is same as the wave function of $d\bar{d}$.
For the wave function of the $s\bar{s}$ components, we also use the
same form as $d\bar{d}$ but with $m^{\eta_{s\bar{s}}}_0$ and $f_s$ instead
of $m^{\eta{d\bar{d}}}_0$ and $f_q$, respectively.

\subsection{Branching ratios}

For the  decays we have considered here, the decay amplitudes in
Eqs.~(\ref{eq:m1}) , (\ref{eq:m2}) and (\ref{eq:m3}) can be
rewritten as
 \beq
 {\cal M} &=& V_{ub}^*V_{us} T -V_{tb}^* V_{ts} P= V_{ub}^*V_{us} T
 \left [ 1 + z e^{ i ( \gamma + \delta ) } \right],
 \label{eq:ma}
 \eeq
 where
 \beq
 z=\left|\frac{V_{tb}^* V_{ts}}{ V_{ub}^*V_{us} } \right|
 \left|\frac{P}{T}\right|
 \label{eq:zz}
 \eeq
 is the ratio of penguin to tree contributions, $\gamma = \arg \left[-\frac{V_{ts}V_{tb}^*}{
 V_{us}V_{ub}^*}\right]$ is the weak phase (one of the three CKM angles), and $\delta$ is the
 relative strong phase
between tree (T) and penguin (P) diagrams.

From Eq.~(\ref{eq:ma}), it is easy to write the decay amplitude for the
corresponding charge conjugated decay mode
 \beq
 \overline{\cal M} &=& V_{ub}V_{us}^* T -V_{tb} V_{ts}^* P
 = V_{ub}V_{us}^* T \left[1 +z e^{i(-\gamma + \delta)} \right].
 \label{eq:mb}
 \eeq
Therefore the CP-averaged branching ratio is
 \beq
 Br = (|{\cal M}|^2 +|\overline{\cal M}|^2)/2 =  \left|
 V_{ub}V_{us}^* T \right| ^2 \left[1 +2 z\cos \gamma \cos \delta
 +z^2 \right], \label{br}
 \eeq
where the ratio $z$ and the strong phase $\delta$ have been defined
in Eqs.(\ref{eq:ma}) and
(\ref{eq:zz}).

Using  the wave functions and the input parameters as specified in
previous sections,  it is straightforward  to calculate the CP
averaged branching ratios for the  considered decays:
\beq
  Br(B_{s} \to \rho^{0} \eta) &=& \left [0.07^{+0.03}_{-0.02}(\omega_{B_s})
    ^{+0.05}_{-0.03}(m_s)^{+0.00}_{-0.00}    (\gamma)\right ]  \times 10^{-6},\non
  Br(B_{s} \to \rho^{0} \eta^{\prime}) &=& \left [0.10^{+0.04}_{-0.03}(\omega_{B_s})
    ^{+0.07}_{-0.04}(m_s)^{+0.00}_{-0.01}
    (\gamma)\right ]  \times 10^{-6},\label{eq:retap}\\
  Br(B_{s} \to \omega \eta) &=& \left [0.21^{+0.04}_{-0.03}(\omega_{B_s})
    ^{+0.14}_{+0.00}(m_s)\pm 0.00
    (\gamma)\right ]  \times 10^{-7},\non
  Br(B_{s} \to \omega \eta^{\prime}) &=& \left [0.13^{+0.04}_{-0.03}(\omega_{B_s})
    ^{+0.05}_{-0.03}(m_s)^{+0.00}_{-0.01}
    (\gamma)\right ]  \times 10^{-6},\label{eq:oetap}\\
  Br(B_{s} \to \phi \eta) &=& \left [2.66^{+1.10}_{-0.74}(\omega_{B_s})
    ^{+1.45}_{-0.78}(m_s)^{+0.01}_{-0.00}
    (\gamma)\right ]  \times 10^{-5},\non
  Br(B_{s} \to \phi \eta^{\prime}) &=& \left [2.00^{+0.80}_{-0.54}(\omega_{B_s})
    ^{+1.42}_{-0.73}(m_s)^{+0.00}_{-0.00}
    (\gamma)\right ]  \times 10^{-5}\label{eq:petap},
   \eeq
where the main errors are induced by the uncertainty of $\omega_{B_s}=0.55 \pm 0.05$ GeV,
$m_s=130 \pm 30$ MeV and $\gamma =60^\circ \pm 20^\circ$, respectively.

As a comparison, we also list here the theoretical predictions based on the QCD factorization approach
as given in Ref.~\cite{jfs03}:
   \beq
  Br(B_{s} \to \rho^{0} \eta) &=& [0.122-0.151] \times 10^{-6},\non
  Br(B_{s} \to \rho^{0} \eta^{\prime}) &=&[ 0.123-0.160]\times 10^{-6},\\
  Br(B_{s} \to \omega \eta) &=& [0.006-0.025] \times 10^{-6},\non
  Br(B_{s} \to \omega \eta^{\prime}) &=& [0.010-0.075] \times 10^{-6},\\
  Br(B_{s} \to \phi \eta) &=& [0.088-0.417]  \times 10^{-6},\non
  Br(B_{s} \to \phi \eta^{\prime}) &=&[ 0.024-0.149] \times 10^{-6}.
   \eeq
One can see that the pQCD predictions are basically consistent with
the corresponding QCDF predictions for the first four decays, but  much larger than the QCDF predictions for
$B_s \to \phi \eta$ and $\phi \eta^\prime$ decays. The difference is about two orders! Further study about
this difference is under way.

 \begin{figure}[tb]
 \vspace{-1cm}
 \centerline{\mbox{\epsfxsize=9cm\epsffile{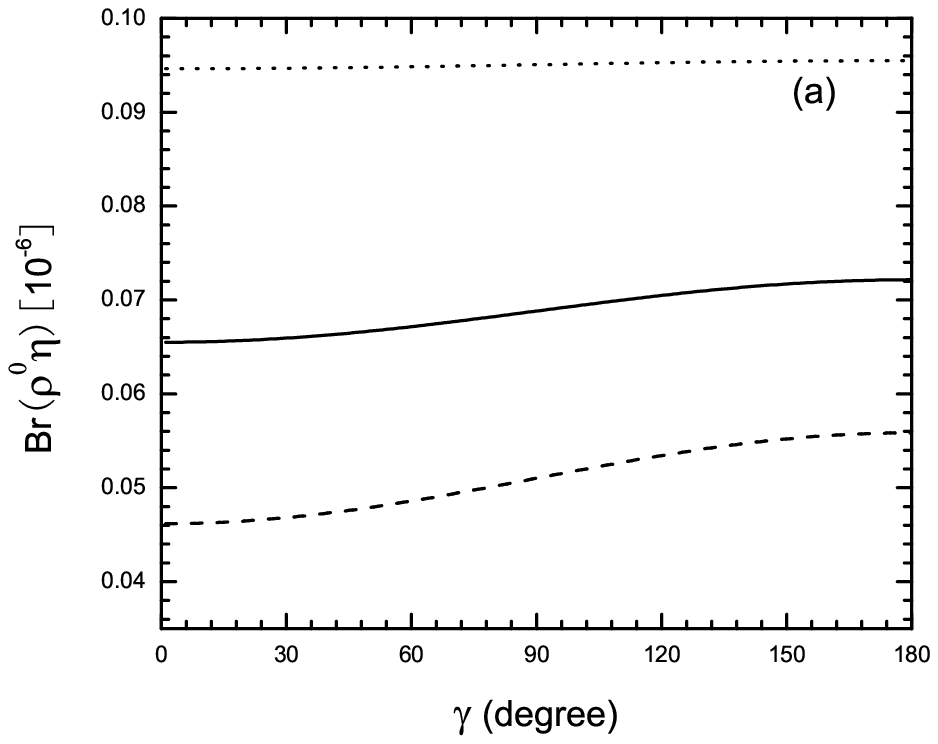}\epsfxsize=9cm\epsffile{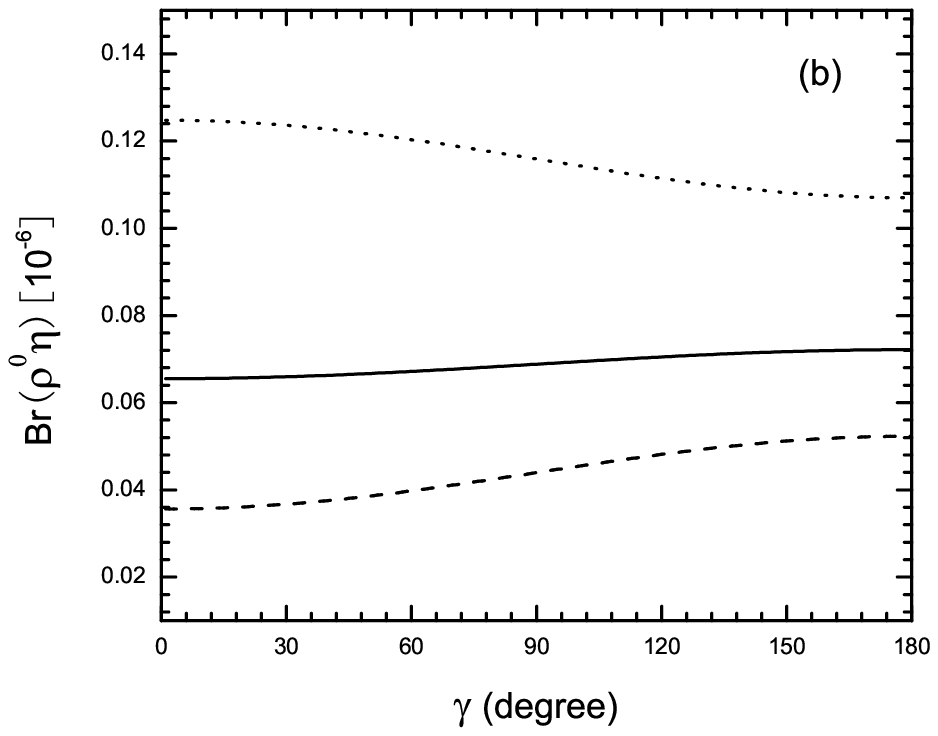}}}
 \vspace{-0.3cm}
 \caption{The CP  averaged branching ratio (in unit of $10^{-6}$) of $B_{s}\to \rho^{0} \eta$
 decay as a function of CKM angle $\gamma$. (a) is for  $m_s=130$ MeV and $\omega_{B_s}=0.50 $ GeV
 (dotted curve), $0.55$ GeV(solid  curve) and $0.60$ GeV(dashed curve); and (b)
 for $\omega_{B_s}=0.55 $ GeV,  and $m_s=100$
 Mev(dotted curve), $130$ MeV (solid  curve) and $160$ MeV (dashed curve).}
  \label{fig:fig2}
 \end{figure}

 \begin{figure}[tb]
\vspace{-1cm}
 \centerline{\mbox{\epsfxsize=9cm\epsffile{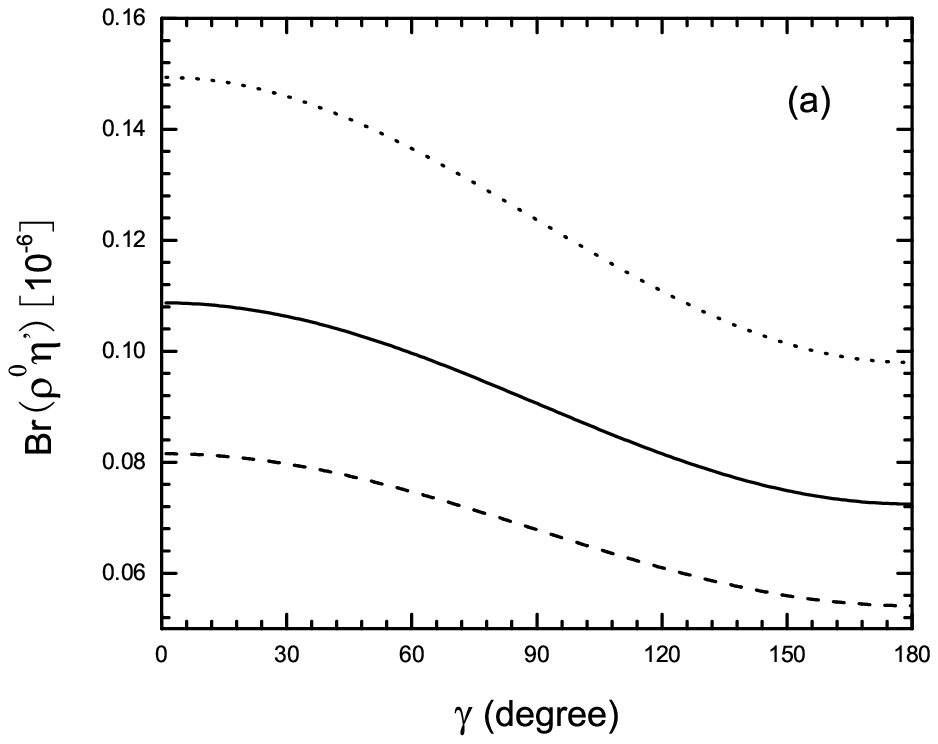} \epsfxsize=9cm\epsffile{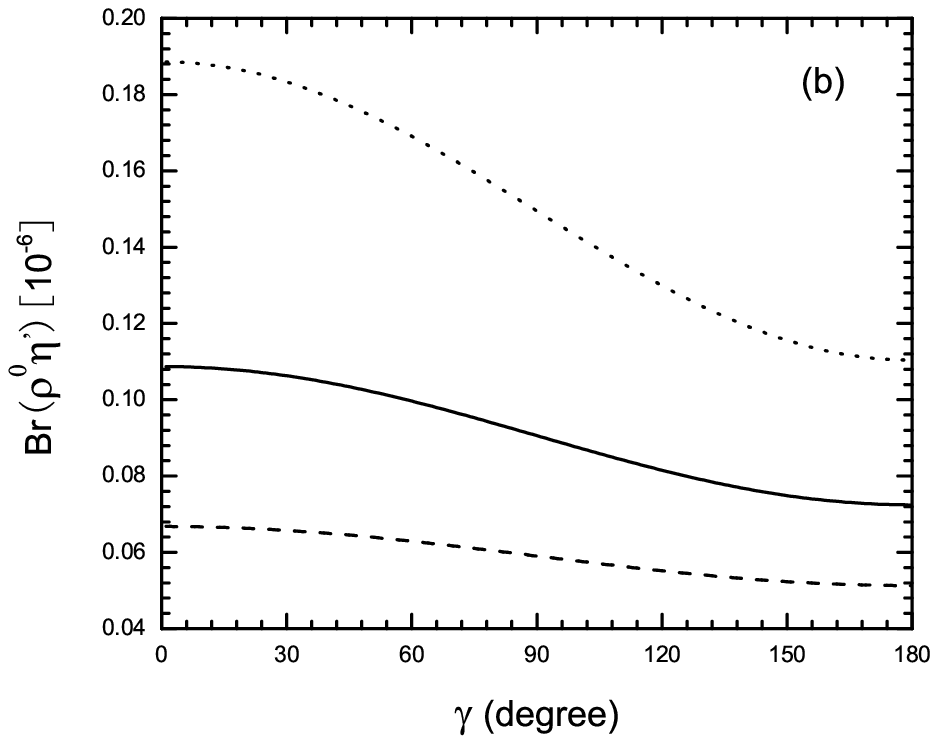}}}
 \vspace{-0.3cm}
 \caption{The same as Fig.~\ref{fig:fig2} but for $B_{s}\to \rho^{0} \eta^{\prime}$  decay. }
  \label{fig:fig3}
 \end{figure}

 \begin{figure}[tb]
\vspace{-1cm}
 \centerline{\mbox{\epsfxsize=9cm\epsffile{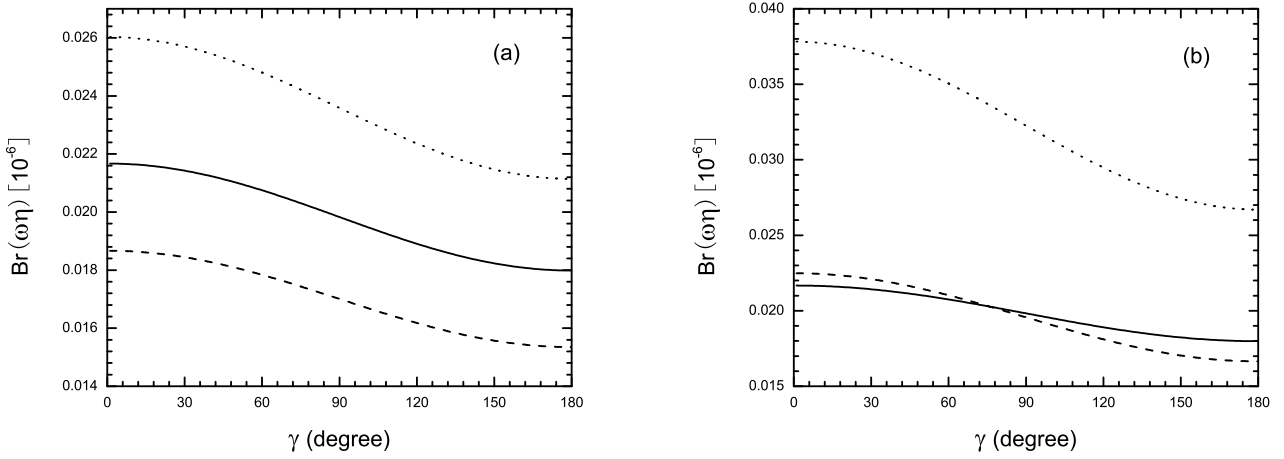} \epsfxsize=9cm\epsffile{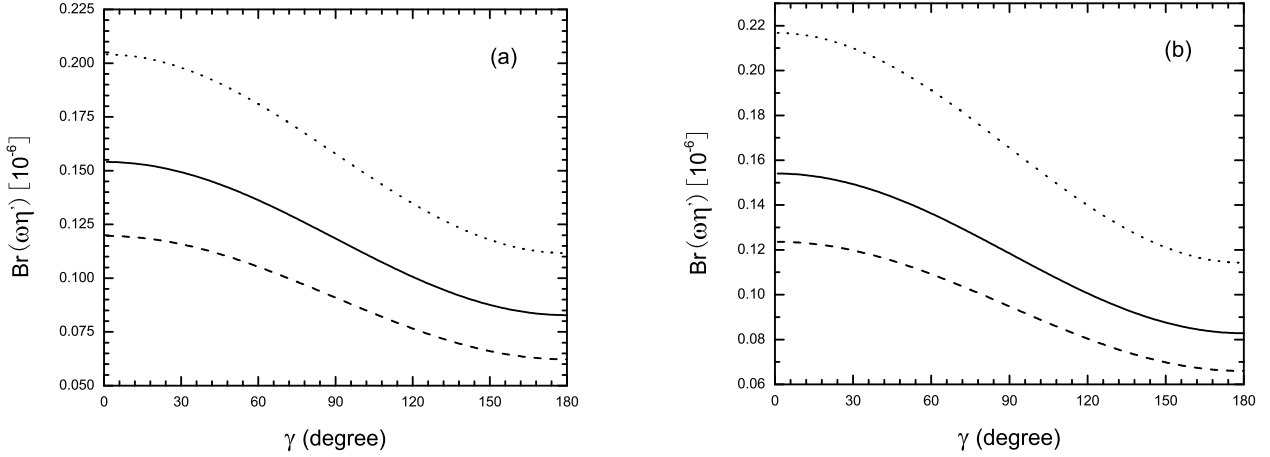}}}
 \vspace{-0.3cm}
 \caption{The same as Fig.~\ref{fig:fig2} but for $B_{s}\to \omega \eta$  decay. }
  \label{fig:fig4}
 \end{figure}

 \begin{figure}[tb]
\vspace{-1cm}
 \centerline{\mbox{\epsfxsize=9cm\epsffile{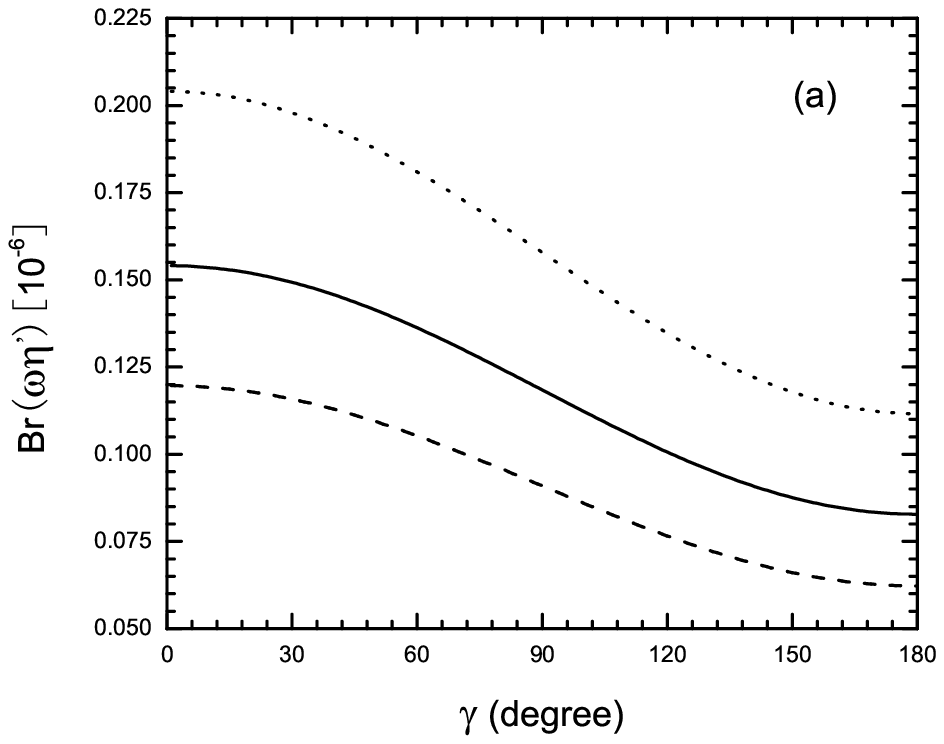}\epsfxsize=9cm\epsffile{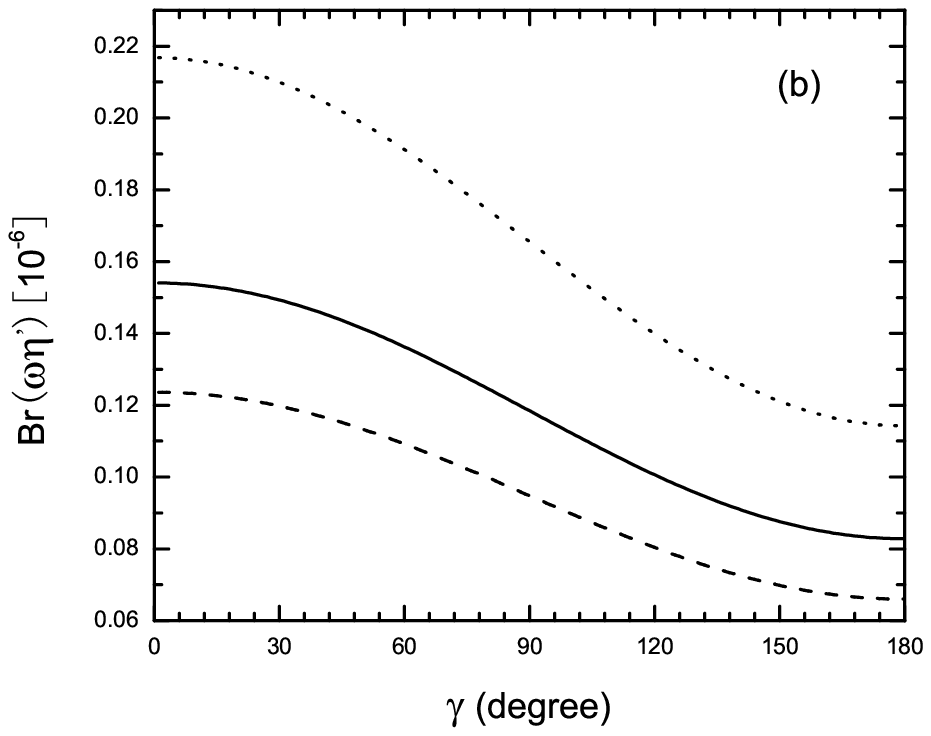}}}
 \vspace{-0.3cm}
 \caption{The same as Fig.~\ref{fig:fig2} but for $B_{s}\to \omega \eta^{\prime}$  decay. }
  \label{fig:fig5}
 \end{figure}

 \begin{figure}[tb]
\vspace{-1cm}
 \centerline{\mbox{\epsfxsize=9cm\epsffile{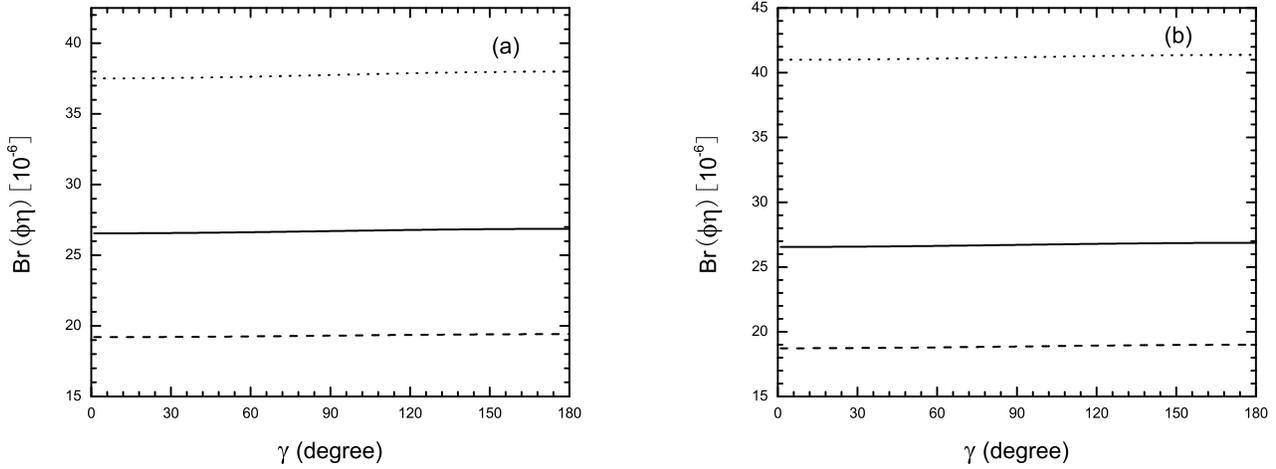} \epsfxsize=9cm\epsffile{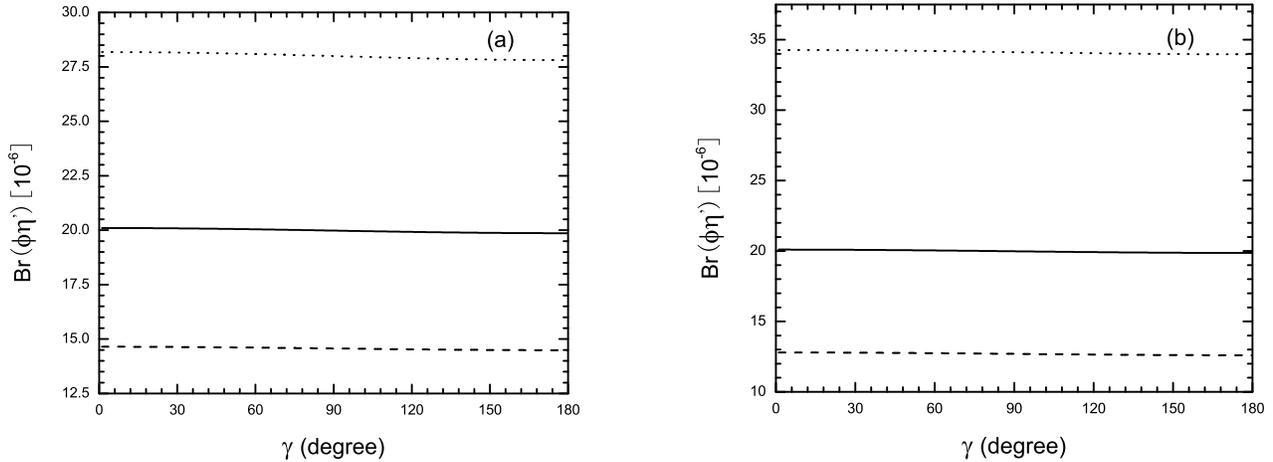}}}
 \vspace{-0.3cm}
 \caption{The same as Fig.~\ref{fig:fig2} but for $B_{s}\to \phi \eta$  decay. }
  \label{fig:fig6}
 \end{figure}

 \begin{figure}[tb]
\vspace{-1cm}
 \centerline{\mbox{\epsfxsize=9cm\epsffile{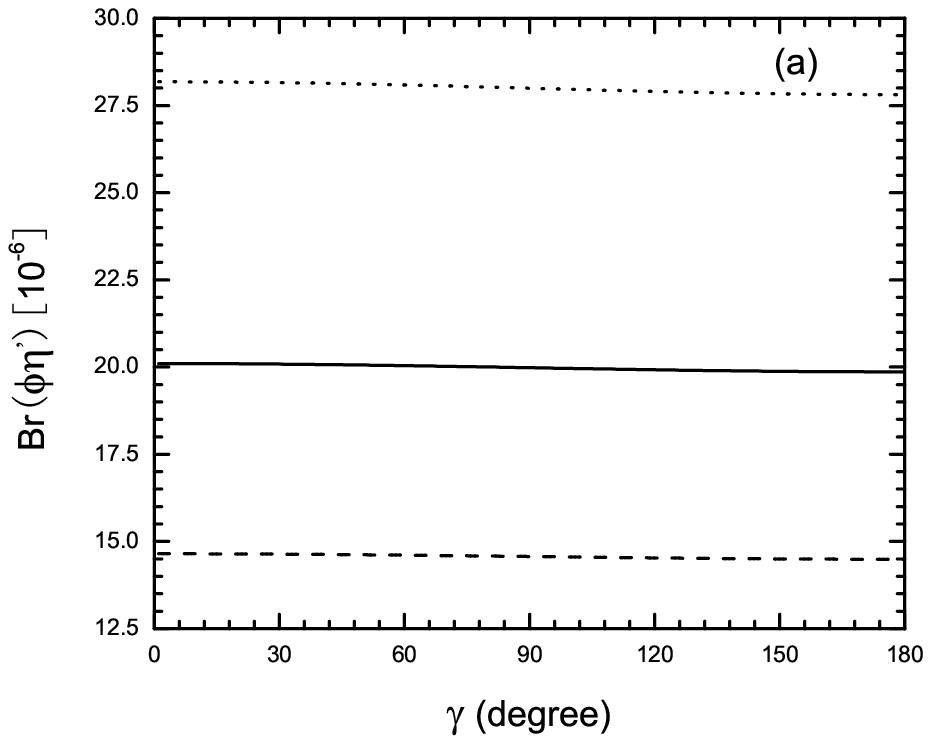} \epsfxsize=9cm\epsffile{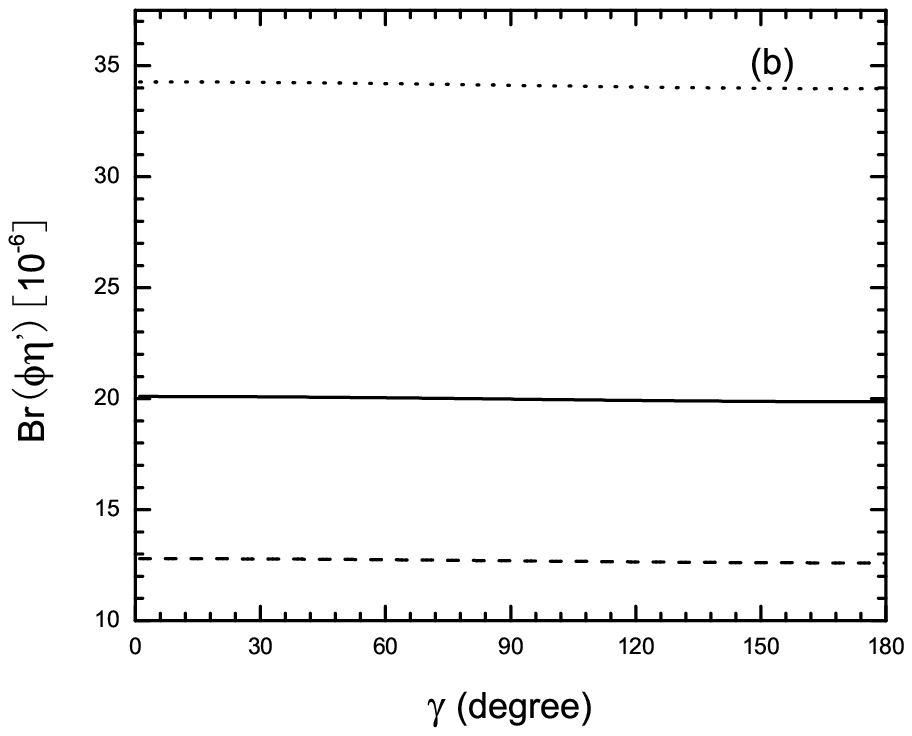}}}
 \vspace{-0.3cm}
 \caption{The same as Fig.~\ref{fig:fig2} but for $B_{s}\to \phi \eta^{\prime}$ decay. }
  \label{fig:fig7}
 \end{figure}

In Figs.~\ref{fig:fig2}-\ref{fig:fig7}, we present the PQCD predictions of
the CP averaged  branching ratios of the considered decays and their
dependence on the variations of $\omega_{B_s}$, $m_s$ and the CKM angle $\gamma$.

\subsection{CP-violating asymmetries }

Now we turn to the evaluation of the CP-violating asymmetries of
the considered decays in PQCD approach. For these neutral decay modes,
the effects of $B_s^0-\bar{B}_s^0$ mixing should be considered.

For $B_s^0$ meson decays, the CP-violating asymmetry of $B_s^0( \bar{B}_s^0) \to f_{CP} $ decay
is time dependent and can be defined as
\beq
A_{CP} &\equiv& \frac{\Gamma\left (\bar{B}_s^0(\Delta t) \to f_{CP}\right)
- \Gamma\left( B_s^0(\Delta t) \to f_{CP}\right )}{
\Gamma\left (\bar{B}_s^0(\Delta t) \to f_{CP}\right )
+ \Gamma\left (B_s^0(\Delta t) \to f_{CP}\right ) }\non
&=& A_{CP}^{dir} \cos (\Delta m_s  \Delta t)
+ A_{CP}^{mix} \sin (\Delta m_s  \Delta t),
\label{eq:acp-def}
\eeq
where $\Delta m_s$ is the mass difference between the two $B_s^0$ mass eigenstates,
$\Delta t =t_{CP}-t_{tag} $ is the time difference between the tagged $B_s^0$
($\bar{B}_s^0$) and the accompanying $\bar{B}_s^0$ ($B_s^0$) with opposite b
flavor decaying to the final CP-eigenstate $f_{CP}$ at the time $t_{CP}$.
The direct and mixing induced CP-violating
asymmetries $A_{CP}^{dir}$ and $A_{CP}^{mix}$ can be written as
\beq
A_{CP}^{dir}=\frac{ \left  | \lambda_{CP}\right |^2 -1 }
{1+|\lambda_{CP}|^2}, \qquad A_{CP}^{mix}=\frac{ 2Im
(\lambda_{CP})}{1+|\lambda_{CP}|^2}, \label{eq:acp-dm}
\eeq
where the CP-violating parameter $\lambda_{CP}$ is
\beq
\lambda_{CP} = \frac{ V_{tb}^*V_{ts} \langle \bar{f}_{CP} |H_{eff}|
 \bar{B}_s^0\rangle} { V_{tb}V_{ts}^* \langle f_{CP}
|H_{eff}| B_s^0\rangle} = e^{2i\gamma}\frac{ 1+z e^{i(\delta-\gamma)} }{
1+ze^{i(\delta+\gamma)} }.
\label{eq:lambda2}
\eeq
Here the ratio $z$ and the strong phase $\delta$ have been defined previously.
In PQCD approach, since both $z$ and $\delta$ are calculable,
it is easy to find the numerical values of
$A_{CP}^{dir}$ and $A_{CP}^{mix}$ for the considered decay processes.

The PQCD predictions for the direct and mixing-induced CP-violating asymmetries are:
  \beq
   A_{CP}^{dir} (B_{s} \to \rho^0 \eta) &=& [4.9^{+0.4}_{-0.1}
   (\omega_{B_s})^{+0.4}_{-0.5}(m_s)^{+0.6}_{-1.2}(\gamma)] \times 10^{-2}, \non
   A_{CP}^{dir} (B_{s} \to \rho^0 \eta^{\prime}) &=& [-25.3^{+1.8}_{-2.0}
   (\omega_{B_s})^{+7.7}_{-8.7}(m_s)^{+7.4}_{-5.3}(\gamma)] \times 10^{-2}, \\
   A_{CP}^{dir} (B_{s} \to \omega \eta) &=& [7.6^{+5.1}_{-5.4}
   (\omega_{B_s})^{+11.1}_{-12.8}(m_s)^{+1.3}_{-2.1}(\gamma)] \times 10^{-2},\non
   A_{CP}^{dir} (B_{s} \to \omega \eta^{\prime}) &=& [-7.6^{+0.4}_{-0.2}
   (\omega_{B_s})^{+3.1}_{-2.8}(m_s)^{+2.3}_{-1.9}(\gamma)] \times 10^{-2}.\\
   A_{CP}^{dir} (B_{s} \to\phi \eta) &=& [0.5^{+0.0}_{-0.1}
   (\omega_{B_s})^{+0.1}_{-0.1}(m_s)^{+0.1}_{-0.1}(\gamma)]\times 10^{-2},\non
   A_{CP}^{dir} (B_{s} \to\phi \eta^{\prime}) &=& [-0.5^{+0.1}_{-0.0}
   (\omega_{B_s})^{+0.1}_{-0.1}(m_s)^{+0.1}_{-0.1}(\gamma)]\times 10^{-2},
   \eeq
  \beq
   A_{CP}^{mix} (B_{s} \to \rho^0 \eta) &=& [-4.2^{+3.8}_{-4.2}
   (\omega_{B_s})^{+10.8}_{-13.1}(m_s)^{+1.1}_{-0.6}(\gamma)] \times 10^{-2}, \non
   A_{CP}^{mix} (B_{s} \to \rho^0 \eta^{\prime}) &=& [18.7^{+0.3}_{+0.5}
   (\omega_{B_s})^{+3.7}_{-3.9}(m_s)^{+1.6}_{-4.4}(\gamma)] \times 10^{-2}, \\
   A_{CP}^{mix} (B_{s} \to \omega \eta) &=& [8.0^{+0.7}_{+0.7}
   (\omega_{B_s})^{+6.4}_{+5.5}(m_s)^{+1.1}_{-2.1}(\gamma)] \times 10^{-2},\non
   A_{CP}^{mix} (B_{s} \to \omega \eta^{\prime}) &=& [24.6^{+1.1}_{-0.6}
   (\omega_{B_s})^{+0.6}_{+0.4}(m_s)^{+4.4}_{-6.8}(\gamma)] \times 10^{-2}.\\
   A_{CP}^{mix} (B_{s} \to\phi \eta) &=& [-0.5^{+0.0}_{-0.0}
   (\omega_{B_s})^{+0.1}_{-0.1}(m_s)^{+0.1}_{-0.1}(\gamma)]\times 10^{-2},\non
   A_{CP}^{mix} (B_{s} \to\phi \eta^{\prime}) &=& [0.5^{+0.1}_{-0.0}
   (\omega_{B_s})^{+0.1}_{-0.1}(m_s)^{+0.1}_{-0.1}(\gamma)]\times 10^{-2},
   \eeq
where the dominant errors come from the variation of $\omega_{B_s}
=0.55\pm 0.05$, $m_s=130\pm 30$ Mev, $\gamma=60^\circ \pm 20^\circ $. One can see from these numerical results
that the CP-violating asymmetries of the considered decay modes are generally not large in size.
This is rather different with the cases of $B_d$ and $B_u$ meson decays, where the CP-violating asymmetries
are generally large in magnitude.

As a simple comparison, we show the QCDF predictions for the direct
CP-violating asymmetries of the six considered decays as given in Ref.~\cite{jfs03}:
\beq
A_{CP}^{dir} (B_{s} \to \rho^0 \eta) &=& -16.9    \times 10^{-2}\non
A_{CP}^{dir} (B_{s} \to \rho^0 \eta^{\prime})&=&   -33.0  \times 10^{-2}, \\
A_{CP}^{dir} (B_{s} \to \omega \eta) &=& 0.76    \times 10^{-2},\non
A_{CP}^{dir} (B_{s} \to \omega \eta^{\prime}) &=& 4.4  \times 10^{-2},\\
A_{CP}^{dir} (B_{s} \to\phi \eta) &=& 23.2  \times 10^{-2},\non
A_{CP}^{dir} (B_{s} \to\phi \eta^{\prime}) &=& -58.3  \times 10^{-2}.
\eeq
Since the relevant measurements are not available at present, and the theoretical uncertainties
in both pQCD and QCFD factorization approaches are still very
large, it is too early to make a meaningful comparison now.
One has to wait for the starting of experimental measurements and improvements of
the theoretical calculations.

In Fig.~\ref{fig:fig6} we show the $\gamma-$dependence of the direct and mixing induced
CP-violating asymmetries for $B_{s}^0 \to \rho^0 \etap$ and $\omega \etap$ decays.
Since the pQCD predictions of the CP asymmetries are sensitive to many input parameters,
the lines in Fig.~\ref{fig:fig6} should be broadened accordingly.

\begin{figure}[tb]
\vspace{-1cm}
 \centerline{\mbox{\epsfxsize=9cm\epsffile{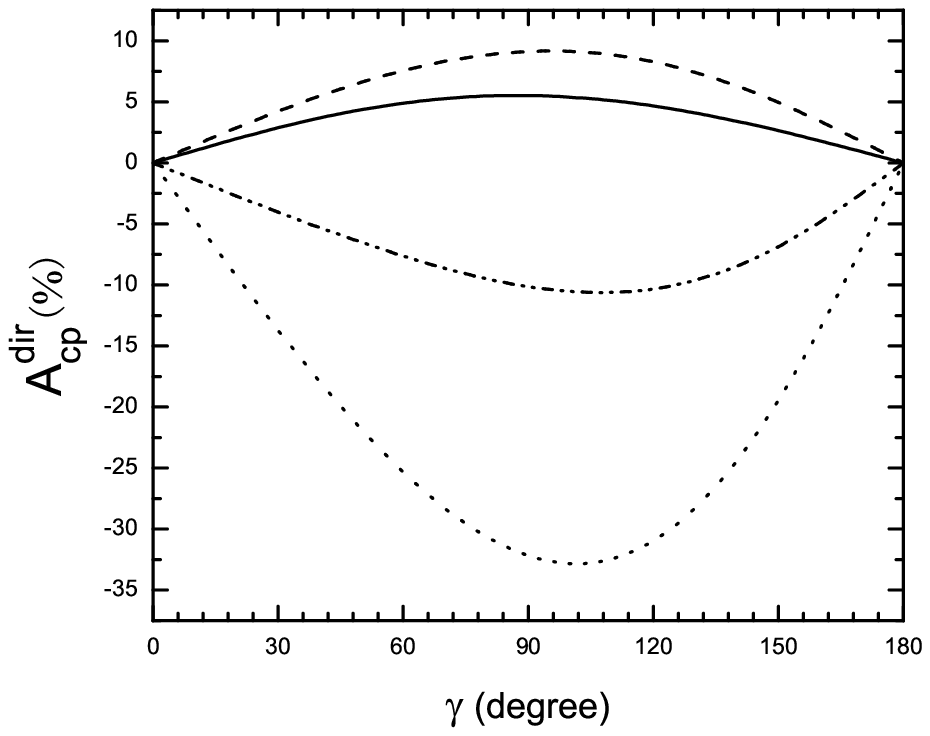} \epsfxsize=9cm\epsffile{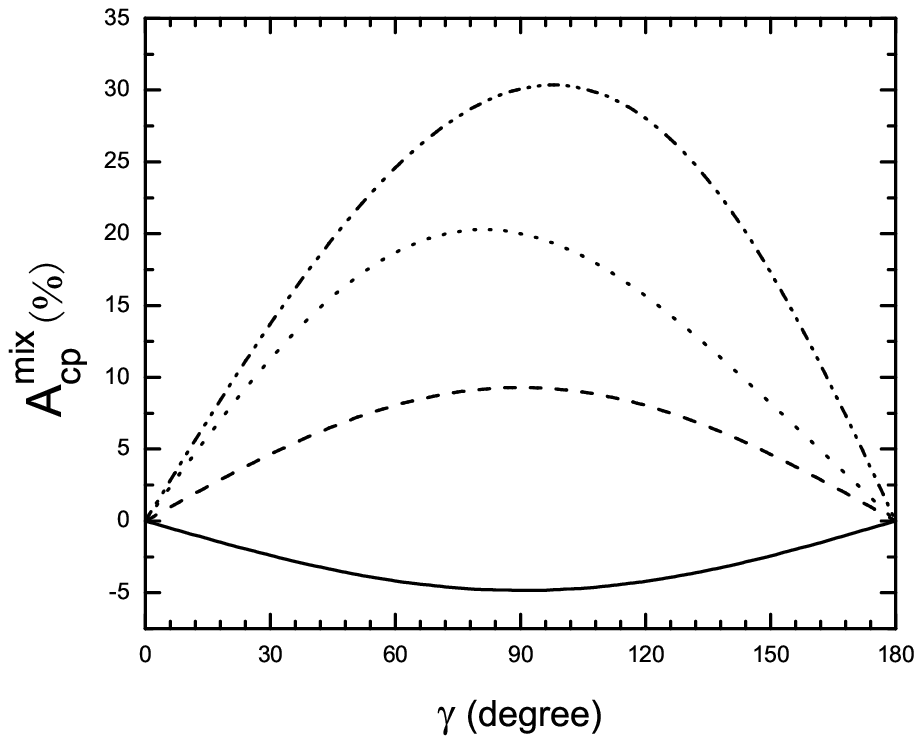}}}
\vspace{-0.3cm}
 \caption{The direct and mixing-induced CP asymmetry (in percentage) of
 $B_s^0 \to \rho^0 \eta$ (solid curve), $\rho^0 \eta^{\prime}$ (dotted curve),
 $\omega \eta$ (dashed-curve), and $\omega \eta^{\prime}$ (dash-dot-dot curve)
 decay as a  function of CKM angle $\gamma$. }
  \label{fig:fig8}
\end{figure}

\subsection{Effects of possible gluonic component of $\eta^\prime$}

Up to now, we have not considered the possible  contributions to the branching
ratios and CP-violating asymmetries of $B \to (\rho^0, \omega,\phi ) \etap$ decays induced by
the possible gluonic component of $\etap$ \cite{ekou2,ek63}.

By using the formulae as given in Ref.~\cite{li0609}, we calculate the gluonic contributions
to $B \to \eta$ and $B \to \eta^\prime$ form factors. For $B \to (\rho, \omega, \phi) \eta$ decays,
the gluonic contributions are negligibly small, less than $3\%$.
For $B \to (\rho, \omega, \phi) \eta^\prime$
decays, the gluonic contributions to the branching ratios are only about $10\%$. The
central values of the pQCD predictions for $B \to (\rho, \omega, \phi) \eta^\prime$ decays
after the inclusion of gluonic contributions are the following
\beq
 Br(B_{s} \to \rho^{0} \eta^{\prime}) &=& \left [0.12^{+0.04}_{-0.03}(\omega_{B_s})
    ^{+0.07}_{-0.04}(m_s)^{+0.00}_{-0.01} (\gamma)\right ]  \times 10^{-6},\\
  Br(B_{s} \to \omega \eta^{\prime}) &=& \left [0.14^{+0.05}_{-0.03}(\omega_{B_s})
    ^{+0.06}_{-0.03}(m_s)^{+0.01}_{-0.01}  (\gamma)\right ]  \times 10^{-6},\\
  Br(B_{s} \to \phi \eta^{\prime}) &=& \left [2.11^{+0.84}_{-0.57}(\omega_{B_s})
   ^{+1.45}_{-0.75}(m_s)^{+0.00}_{-0.00} (\gamma)\right ]  \times 10^{-5}.
\eeq

Again, the gluonic contributions are small in size, which is consistent with previous
results for other B meson decays involving $\eta$ or $\eta^\prime$ as one or two final state
mesons \cite{liu06,wang06,xiao0606,xiao07}.

\section{summary }

In this paper, we calculated the branching ratios and CP-violating
asymmetries of $B_{s} \to \rho^0 \etap$, $B_{s} \to\omega \etap$, and $B_{s}\to \phi \etap$ decays in
the pQCD factorization approach.

Besides the usual factorizable diagrams, the non-factorizable and
annihilation diagrams as shown in Fig.~(\ref{fig:fig1}) are also
calculated analytically. Although the non-factorizable and
annihilation contributions are sub-leading for the branching ratios
of the considered decays, but they are not negligible. Furthermore
these diagrams provide the necessary strong phase required for a
non-zero CP-violating asymmetry for the considered decays.

From our calculations and phenomenological analysis, we found the following results:
\begin{itemize}

\item
For the branching ratios of the six considered decay modes, the pQCD predictions are:
   \beq
  Br(B_{s} \to \rho^{0} \eta) &=& [0.07^{+0.06}_{-0.04}] \times 10^{-6},\non
  Br(B_{s} \to \rho^{0} \eta^{\prime}) &=&[ 0.10^{+0.08}_{-0.05}]\times 10^{-6},\\
  Br(B_{s} \to \omega \eta) &=& [0.02^{+0.015}_{-0.003}] \times 10^{-6},\non
  Br(B_{s} \to \omega \eta^{\prime}) &=& [0.13^{+0.06}_{-0.04}] \times 10^{-6},\\
  Br(B_{s} \to \phi \eta) &=& [2.66^{+1.82}_{-1.08}]  \times 10^{-5},\non
  Br(B_{s} \to \phi \eta^{\prime}) &=&[ 2.00^{+1.63}_{-0.91}] \times 10^{-5}.
 \eeq
Here the various errors as specified previously have been
added in quadrature. The pQCD predictions for the first four decay channels agree well
with those obtained by employing the QCDF approach.
For last two decays, however, there are large differences between the pQCD and QCDF approach,
which will be tested in the forthcoming LHC-b experiments.

\item
The gluonic contributions are small in size: less than $3\%$ for $B \to (\rho, \omega, \phi) \eta$
decays, and about $10\%$ for $B \to (\rho, \omega, \phi) \eta^\prime$ decays.

\item
For the CP-violating asymmetries, the pQCD predictions are generally not large in size,
but still have large theoretical uncertainties.

 \end{itemize}

\begin{acknowledgments}
We are very grateful to Cai-Dian L\"u, Hui-sheng Wang , Xin Liu and Ying Li for
helpful discussions.
This work is partly supported  by the National Natural Science
Foundation of China under Grant No.10575052, by the
Specialized Research Fund for the doctoral Program of higher education (SRFDP)
under Grant No.~20050319008 and by the Research Fund from Nanjing Normal University under
Grant No.~214080A916.

\end{acknowledgments}


\begin{appendix}

\section{Related Functions }\label{sec:aa}

We show here the function $h_i$'s, coming from the Fourier transformations  of $H^{(0)}$,
 \beq
 h_e(x_1,x_3,b_1,b_3)&=&
 K_{0}\left(\sqrt{x_1 x_3} m_{B_{s}} b_1\right)
 \left[\theta(b_1-b_3)K_0\left(\sqrt{x_3} m_{B_{s}}
b_1\right)I_0\left(\sqrt{x_3} m_{B_{s}}b_3\right)\right.
 \non
& &\;\left. +\theta(b_3-b_1)K_0\left(\sqrt{x_3}  m_{B_{s}}
b_3\right) I_0\left(\sqrt{x_3}  m_{B_{s}} b_1\right)\right]
S_t(x_3), \label{he1}
 \eeq
 \beq
 h_a(x_2,x_3,b_2,b_3)&=&
 K_{0}\left(i \sqrt{x_2 x_3} m_{B_{s}} b_2\right)
 \left[\theta(b_3-b_2)K_0\left(i \sqrt{x_3} m_{B_{s}}
b_3\right)I_0\left(i \sqrt{x_3} m_{B_{s}} b_2\right)\right.
 \non
& &\;\;\;\;\left. +\theta(b_2-b_3)K_0\left(i \sqrt{x_3}  m_{B_{s}}
b_2\right) I_0\left(i \sqrt{x_3}  m_{B_{s}} b_3\right)\right]
S_t(x_3), \label{he3} \eeq
 \beq
 h_{f}(x_1,x_2,x_3,b_1,b_2) &=&
 \biggl\{\theta(b_2-b_1) \mathrm{I}_0(M_{B_{s}}\sqrt{x_1 x_3} b_1)
 \mathrm{K}_0(M_{B_{s}}\sqrt{x_1 x_3} b_2)
 \non
&+ & (b_1 \leftrightarrow b_2) \biggr\}  \cdot\left(
\begin{matrix}
 \mathrm{K}_0(M_{B_{s}} F_{(1)} b_2), & \text{for}\quad F^2_{(1)}>0 \\
 \frac{\pi i}{2} \mathrm{H}_0^{(1)}(M_{B_{s}}\sqrt{|F^2_{(1)}|}\ b_2) ,&
 \text{for}\quad F^2_{(1)}<0
\end{matrix}
\right),
\label{eq:pp1}
 \eeq
 \beq
 h_f^1(x_1,x_2,x_3,b_1,b_2) &=&
 \biggl\{\theta(b_1-b_2) \mathrm{K}_0(i \sqrt{x_2 x_3} b_1 M_{B_{s}})
 \mathrm{I}_0(i \sqrt{x_2 x_3} b_2 M_{B_{s}})
 \non
&+& (b_1 \leftrightarrow b_2) \biggr\} \cdot \left(
\begin{matrix}
 \mathrm{K}_0(M_{B_{s}} F_{(2)} b_1), & \text{for}\quad F^2_{(2)}>0 \\
 \frac{\pi i}{2} \mathrm{H}_0^{(1)}(M_{B_{s}}\sqrt{|F^2_{(2)}|}\ b_1), &
 \text{for}\quad F^2_{(2)}<0
\end{matrix}\right),
\label{eq:pp3}
 \eeq
 \beq
 h_f^2(x_1,x_2,x_3,b_1,b_2) &=&
\biggl\{\theta(b_1-b_2) \mathrm{K}_0(i \sqrt{x_2 x_3} b_1 M_{B_{s}})
 \mathrm{I}_0(i \sqrt{x_2 x_3} b_2 M_{B_{s}})
  \non
 &+& (b_1 \leftrightarrow b_2) \biggr\} \cdot \left(
 \begin{matrix}
 \mathrm{K}_0(M_{B_{s}} F_{(3)} b_1), & \text{for}\quad F^2_{(3)}>0 \\
 \frac{\pi i}{2} \mathrm{H}_0^{(1)}(M_{B_{s}}\sqrt{|F^2_{(3)}|}\ b_1), &
 \text{for}\quad F^2_{(3)}<0
 \end{matrix}
 \right),
 \label{eq:pp4}
 \eeq
where $J_0$ is the Bessel function,   $K_0$ and  $I_0$ are modified
Bessel functions $K_0 (-i x) = -(\pi/2) Y_0 (x) + i (\pi/2) J_0
(x)$; $\mathrm{H}_0^{(1)}(z)$ is the Hankel function,
$\mathrm{H}_0^{(1)}(z) = \mathrm{J}_0(z) + i\, \mathrm{Y}_0(z)$, and $F_{(j)}$'s are defined by
 \beq
F^2_{(1)}&=&(x_1 -x_2) x_3\;,\\
F^2_{(2)}&=&(x_1-x_2) x_3\;,\\
F^2_{(3)}&=& x_1+x_2+x_3-x_1 x_3-x_2 x_3 \;\;.
 \eeq

The threshold resummation form factor $S_t(x_i)$ is adopted from Ref.\cite{tk07}
\beq
S_t(x)=\frac{2^{1+2c} \Gamma
(3/2+c)}{\sqrt{\pi} \Gamma(1+c)}[x(1-x)]^c,
\eeq
where the parameter $c=0.3$. This function is normalized to unity. More
information about the threshold resummation can be found in
reference \cite{hnl53,hnl66}.

The Sudakov factors used in the text are defined as
 \beq
S_{ab}(t) &=& s\left(x_1 m_{B_{s}}/\sqrt{2}, b_1\right) +s\left(x_3
m_{B_{s}}/\sqrt{2}, b_3\right) +s\left((1-x_3) m_{B_{s}}/\sqrt{2},
b_3\right) \non
&&-\frac{1}{\beta_1}\left[\ln\frac{\ln(t/\Lambda)}{-\ln(b_1\Lambda)}
+\ln\frac{\ln(t/\Lambda)}{-\ln(b_3\Lambda)}\right],
\label{wp}\\
S_{cd}(t) &=& s\left(x_1 m_{B_{s}}/\sqrt{2}, b_1\right)
 +s\left(x_2 m_{B_{s}}/\sqrt{2}, b_2\right)
+s\left((1-x_2) m_{B_{s}}/\sqrt{2}, b_2\right) \non
 && +s\left(x_3
m_{B_{s}}/\sqrt{2}, b_1\right) +s\left((1-x_3) m_{B_{s}}/\sqrt{2},
b_1\right) \non
 & &-\frac{1}{\beta_1}\left[2
\ln\frac{\ln(t/\Lambda)}{-\ln(b_1\Lambda)}
+\ln\frac{\ln(t/\Lambda)}{-\ln(b_2\Lambda)}\right],
\label{Sc}\\
S_{ef}(t) &=& s\left(x_1 m_{B_{s}}/\sqrt{2}, b_1\right)
 +s\left(x_2 m_{B_{s}}/\sqrt{2}, b_2\right)
+s\left((1-x_2) m_{B_{s}}/\sqrt{2}, b_2\right) \non
 && +s\left(x_3
m_{B_{s}}/\sqrt{2}, b_2\right) +s\left((1-x_3) m_{B_{s}}/\sqrt{2},
b_2\right) \non
 &&-\frac{1}{\beta_1}\left[\ln\frac{\ln(t/\Lambda)}{-\ln(b_1\Lambda)}
+2\ln\frac{\ln(t/\Lambda)}{-\ln(b_2\Lambda)}\right],
\label{Se}\\
S_{gh}(t) &=& s\left(x_2 m_{B_{s}}/\sqrt{2}, b_2\right)
 +s\left(x_3 m_{B_{s}}/\sqrt{2}, b_3\right)
+s\left((1-x_2) m_{B_{s}}/\sqrt{2}, b_2\right) \non
 &+& s\left((1-x_3)
m_{B_{s}}/\sqrt{2}, b_3\right)
-\frac{1}{\beta_1}\left[\ln\frac{\ln(t/\Lambda)}{-\ln(b_3\Lambda)}
+\ln\frac{\ln(t/\Lambda)}{-\ln(b_2\Lambda)}\right], \label{ww}
 \eeq
where the function $S_{B_s}$, $S_{\rho^0}$, $S_{\etap}$ used in
the amplitudes are defined as:
 \beq
 S_{B_s}(t) &=& s(x_1P_1^+,b_1)+2\int_{1/b_1}^t\!\!\!\frac{d\bar\mu}{\bar\mu}
 \gamma(\alpha_s(\bar\mu)),\\
  S_{\rho^0}(t) &=& s(x_2P_2^+,b_2)+s\left((1-x_2)P_2^+,b_2\right)+
 2\int_{1/b_2}^t\!\!\!\frac{d\bar\mu}{\bar\mu}\gamma\left(\alpha_s(\bar\mu)\right),\\
 S_\etap(t) &=& s(x_3P_3^-,b_3)+s\left((1-x_3)P_3^-,b_3\right)+
 2\int_{1/b_3}^t\!\!\!\frac{d\bar\mu}{\bar\mu}\gamma\left(\alpha_s(\bar\mu)\right).
 \eeq
where the so called Sudakov factor $s(Q,b)$ resulting from the
resummation of double logarithms is given as:
 \beq
s(Q,b)=\int_{1/b}^Q\!\!\! \frac{d\mu}{\mu}\Bigl[
\ln\left(\frac{Q}{\mu}\right)A(\alpha(\bar\mu))+B(\alpha_s(\bar\mu))
\Bigr] \label{su1}
 \eeq
with
\begin{gather}
A=C_F\frac{\alpha_s}{\pi}+\left[\frac{67}{9}-\frac{\pi^2}{3}-\frac{10}{27}n_{f}+
\frac{2}{3}\beta_0\ln\left(\frac{e^{\gamma_E}}{2}\right)\right]
 \left(\frac{\alpha_s}{\pi}\right)^2 ,\\
B=\frac{2}{3}\frac{\alpha_s}{\pi}\ln\left(\frac{e^{2\gamma_{E}-1}}{2}\right).\hspace{6cm}
\end{gather}
Here $\gamma_E=0.57722\cdots$ is the Euler constant, $n_{f}$ is the
active quark flavor number. For the detailed derivation of the Sudakov factors, see Ref.\cite{hnl88}.
The hard scale $t_i$'s in the above equations are chosen as:
\beq
t_{e}^1 &=& {\rm max}(\sqrt{x_3} m_{B_{s}},1/b_1,1/b_3)\;,\\
t_{e}^2 &=& {\rm max}(\sqrt{x_1}m_{B_{s}},1/b_1,1/b_3)\;,\\
t_{e}^3 &=& {\rm max}(\sqrt{x_3}m_{B_{s}},1/b_2,1/b_3)\;,\\
t_{e}^4 &=& {\rm max}(\sqrt{x_2}m_{B_{s}},1/b_2,1/b_3)\;,\\
t_{f} &=& {\rm max}(\sqrt{x_1 x_3}m_{B_{s}},
\sqrt{x_2 x_3} m_{B_{s}},1/b_1,1/b_2)\;,\\
t_{f}^1 &=& {\rm max}(\sqrt{x_2x_3} m_{B_{s}},1/b_1,1/b_2)\;,\\
 t_{f}^2 &=& {\rm max}(\sqrt{x_1+x_2+x_3-x_1 x_3-x_2 x_3}m_{B_{s}}, \sqrt{x_2 x_3} m_{B_{s}},1/b_1,1/b_2)\;.
\eeq
They are given as the maximum energy scale appearing in each diagram
to kill the large logarithmic radiative corrections.

\end{appendix}


\end{document}